\documentclass[a4paper, 11pt]{article}

\usepackage{doublespace}
\usepackage{astron}
\usepackage{amsmath,amssymb}
\usepackage{palatino}
\usepackage{graphicx, subfigure}
\def\etal{et~al.}
\def\2mass{2MASS}

\def\tf{Tully-Fisher}
\def\Gio{Giovanelli \etal}
\def\Gios{Giovanelli {\etal}s'}
\def\in{\textbf{in}}
\def\inp{\textbf{in+}}

\def\wrt{with respect to}
\def\deg{$^\circ$}
\def\Wcor{W_{\text{cor}}}
\def\scr{\scriptsize}

\title{\Huge The 2MASS \tf\ Relation\\ and Local Peculiar Velocities}

\author{{\LARGE Steven P. Bamford}\\
\begin{minipage}{350pt}
        \center
        \medskip
        {\small steven.bamford@physics.org}\\
        \bigskip 
        \textit{\Large Final Report on Level~4 Project}\\
        \bigskip
        MSci Physics and Astronomy\\\medskip
        University of Durham\\
        \bigskip
        \textit{May~2002}
\end{minipage}}
\date{}

\begin{document}

\begin{singlespace}
\maketitle
\end{singlespace}

\begin{abstract}
This study assesses the utility of applying the Tully-Fisher (TF) relation using photometric data
from the 2MASS project.  This is acheived by performing a reasonably detailed preliminary analysis
using 2MASS extended source data with the SCI sample of \Gio\ \cite*{Gio97_data}.
Distances and peculiar velocities are measured for 11 clusters out to $\sim$$75$ $h^{-1}$ Mpc.
Statistics are found to be limited by the 47\% coverage of the current 2MASS second incremental release.

The 2MASS $J$, $H$ and $K$-band photometry produce results
in good agreement with those derived from the SCI $I$-band data.
These results are also generally in reasonable agreement with \Gios\ \cite*{Gio97_analysis,G98cluster} own analysis.
Comparison of the results obtained with two different 2MASS magnitudes suggests they are not
overly sensitive to the type of magnitude used.
It is concluded that the 2MASS TF is a useful tool for distance and peculiar velocity work, especially
with the improved coverage that will be provided by the full 2MASS release.

A detailed discussion of the Tully-Fisher relation and previous work, particularly that in the near-infra-red,
is presented.  2MASS and its extended source catalogue are described, and the best magnitudes for TF work considered.
A number of corrections to the photometry are discussed and applied.
The SCI sample is described, and the effect on statistics of combining this sample with 
the 2MASS second incremental release is assessed. 
The problems encountered when fitting the TF have
been investigated, and a method for avoiding various biases has been implemented.
\end{abstract}
\newpage

\begin{singlespace}\tableofcontents\end{singlespace}

\newpage
\pagestyle{myheadings}
\markright{The 2MASS \tf\ Relation and Local Peculiar Velocities}

\section{Introduction}
\subsection{The \tf\ relation} \label{sec:tf}

There is a  general correlation between the luminosity of a spiral galaxy and its rotational velocity,
first noticed by a number of investigators in the early 1970s (e.g. Balkowski \etal\ \cite*{BBCGH74},
Rogstad \& Shostak \cite*{RS72}, Shostak \cite*{Shostak75}). 
The first detailed application of this correlation was by Tully \& Fisher \cite*{TF76,TF77},
and therefore it is commonly referred to as the \tf\ relation.
Their investigation applied the correlation between apparent visible magnitude (or photometric diameter) 
and HI line-width (a measure of a galaxy's rotational velocity) to derive distances to the Virgo and Ursa Major clusters
and estimate a value of $80$ km s$^{-1}$ Mpc$^{-1}$ for the Hubble constant.

The manner in which the \tf\ relation has been applied since varies between studies,
and is a long standing point of debate.
All approaches use essentially the same correlation, but it is the manner in which the two principle observed 
quantities --- luminosity and circular velocity --- are measured, corrected and analysed that gives
variations in the technique.  As part of the correction procedure a measurement of inclination and/or axial ratio
is also required.  The quality and manner in which this is obtained may affect the results.

Different observational quantities may be employed to represent the physical variables.
For example, the circular velocity has usually been characterised by a measurement of the 
HI line-width, while a number of modern studies use optical, long-slit spectroscopy to derive a
rotation curve (eg. Dale \etal\ \cite*{Dale99_V}).
There are also many different definitions of magnitude,
which may be applied directly or used to determine a total luminosity which is subsequently applied in the relation.
Many studies, in an effort to maximise the
number of galaxies in their sample, combine data from a number of sources. They compare the overlapping
objects and attempt to homogenise the disparate data with source dependent corrections.
Examples of this approach include the SCI sample of \Gio\ \cite*{Gio97_data},
and the Mark III catalogue of Willick and collaborators \cite{MarkIII-III},
and their analyses \cite{Gio97_analysis,MarkIII-I,MarkIII-II}.

The photometry may be measured in different spectral bands.
In particular there are advantages to working in the infra-red (IR).  These were highlighted by
Aaronson and co-workers shortly after the TF relation was proposed \cite{AM78,AMH79}.
The absorption of light by dust is significantly lower at longer wavelengths.
The extinctions, in both our Galaxy and internal to the observed galaxy, are thus reduced,
and correspondingly so are the corrections to the photometry required to account for these effects.
Therefore the additional uncertainties on magnitude that are contributed by these corrections
are less significant. The results are not as sensitive to the details of the corrections used and 
are hence more robust.
Additional to this is the ability to observe closer to the Galactic plane, where dust obscuration
prohibits optical work.

However, the principle TF-specific advantage extolled by Aaronson \etal, is that galaxies are intrinsically red objects.
Even in spirals, which may appear blue overall, the kinematically representative stellar population
comprises the long-lived, low-mass stars, whose spectra peak in the near-infra-red (NIR).
Being older and more evenly-distributed and numerous, these stars better represent the long-term average
properties of a galaxy than the young, massive, blue population.  Thus working in the IR dampens the 
short-term variations in galaxy luminosity, and hence reduces the scatter these would add to the TF relation.

An additional practical benefit of working in the IR is that the smooth distribution of the redder population produces
an image with simple, elliptical isophotes.  These are easier to fit for adaptive aperture
photometry than the spiral arms seen in the visible, and hence give more reliable magnitudes.

It was immediately realised \cite{TF77,ST76} that corrections to the photometry for extinction,
both Galactic and internal to the galaxy under study, are required to remove systematic errors.
These corrections vary in form as the analyses become more sophisticated. 
Additional dependencies of the relation, such as on morphological type have been proposed,
suggesting further possibilities for correction.
Other features of the technique, such as intrinsic scatter and selection biases, have been noticed,
and a number of methods are used to account for these in fitting the relation.
These corrections, biases and fitting procedures are discussed later in this report, in the context of my analysis.

Virtually all analyses differ to some extent in at least one of; the measured quantities, corrections or
fitting methods.
This variety of methodology complicates direct comparison of the TF parameters between studies.
Comparisons must rather be made between quantities and conclusions derived from the TF,
which are ideally independent of the method, such as distances.

The TF relation is one of the most important tools in extragalactic distance estimation, 
and is probably \emph{the} best for spiral galaxies at distances beyond our local group.
This importance is demonstrated by the fact that it is the focus of a Hubble Space Telescope Key Project
\cite{Sakai_etal}.  With the HST's resolution it is possible to identify Cepheid variables in a larger number 
of nearby galaxies than was previously possible, and in particular galaxies which can be used with the TF.
This enables a direct calibration of the TF relation, and drastically reduces the number of steps, and hence
the uncertainty, in the extragalactic distance scale.

There are a number of alternative extragalactic distance estimators.
These all utilize some relation between a distance-independent observable and absolute magnitude
The Type Ia supernovae technique \cite{SNIa96} is effective to large distances, but suffers 
from sparse coverage due to its serendipitous nature.
Recent work on surface brightness fluctuations shows promise \cite{FP-SBF-02}, 
but its reach is limited in comparison with the TF.
The Fundamental Plane \cite{FP}, 
and $D_n - \sigma$ \cite{Dn-sigma} methods are both evolved from the original Faber-Jackson relation \cite{Faber-Jackson}.
However these are specific to elliptical galaxies, which are only usually found
in clusters.  Indeed, current theories suggest elliptical galaxies are formed from spirals as they
fall into the harsh cluster environment \cite{MM98,KS01}.  This process has an advantage in that it 
removes the influence of a galaxy's initial formation on it's current properties,
and ellipticals are actually seen to have low individual variation
by the small scatter on these relations \cite{FP-SBF-02}.  However, there are substantial differences between
individual cluster environments and their evolution with redshift is uncertain.  These variations may 
give rise to differences in the elliptical populations of each cluster which are hard to determine
independently, and hence correct for.  In contrast, spiral galaxies are found both in the field and associated with
clusters.  However, spirals in clusters tend to be located in their outer regions, meaning they are mostly unaffected
by the cluster environment.  There is thus less uncertainty on the TF relation 
from environmental and evolutionary effects.

\subsection{Physical origins of the \tf}%
\footnote{this section generally uses information from Rhee \cite*{Rhee96}.}
\label{sec:physical_origins}
In addition to a purely practical distance estimator, the existence of the TF relation is important 
for astrophysical considerations.
The observed TF parameters provide insights into the dynamics of galaxies, and impose constraints on 
theories of galaxy formation and evolution.
Modern, large, numerical simulations of galaxies do reproduce a TF relation.
For example, Koda, Sofue \& Wada \cite*{Koda_etal} successfully reproduce a general \tf\ with a comparable slope and scatter,
identifying the slope corresponding to varying galaxy masses, and the intrinsic scatter due to variations in the initial spin.
However, the details of formation, and effects of external forces, are not considered.

It is obvious that there must be some more to the TF relation than just a correlation between luminosity 
and circular velocity.  The two cannot be \emph{directly} physically related.  It is generally thought that there are a 
number of independent sub-relations forming the, more easily observed, overall TF.  These are primarily relations between,
\begin{singlespace} \vspace{-2\parskip}
\begin{itemize}
\item total luminosity and luminous mass (characterised by the star formation history and initial mass function),
\item luminous mass and total mass distribution (including dark matter),
\item total mass distribution and outer circular velocity (determined by the galaxy formation process).
\end{itemize}
\end{singlespace}
Each of these sub-relations could, in theory, be used as an independent distance estimator.
However, the luminosity and rotational velocity are the most easily, and directly, measured quantities in the above list.
Estimates of galaxy mass are either very approximate (often maximum limits), 
and/or rely heavily on theories of galaxy formation.  
They are also virtually always derived from measurements of rotational velocity or luminosity profile, 
and often both of these \cite{Rhee96}, and hence require a detailed bias consideration.

\subsection{Previous Work} \label{sec:previous}

For all the reasons stated in the previous section, the TF relation is extensively used, 
and has been applied with varying success in determinations of distances, $H_0$ and peculiar velocities 
(additional to the Hubble flow).
It is also useful for putting constraints on
theories of galaxy formation and evolution (\S \ref{sec:physical_origins}).

The original analysis of Tully \& Fisher \cite*{TF77}, considering local galaxies and the Virgo and Ursa Major
clusters, produced an estimate of $H_0 = 80$ km s$^{-1}$ Mpc$^{-1}$.  Using the same method with a slightly
augmented sample, though not including Ursa Major, and with some data from other sources,
Sandage \& Tammann \cite*{ST76} found $H_0 = 50$ km s$^{-1}$ Mpc$^{-1}$.
The causes of this serious disagreement may be attributed to a number of factors, but in particular 
to the inclusion of more galaxies closer to face on orientation combined with a linear internal 
extinction correction (as opposed to the preferred logarithmic correction now in use), and also the
effect of not including Ursa Major.

Following their original advocation of the IR TF and the preliminary analyses of this cited above,
Aaronson, Mould, Huchra and collaborators embarked upon a detailed study of the TF distance scale
, and it's implications for the local cosmology
\cite[hereafter A80I, A80II, A80III respectively]{AHM80,MAH80,AMHSSB80}.
Calibrating with local galaxies, using a variety of other distance indicators (A80I), and then applying the relation
to the Virgo cluster they derived a value of $H_0 = 65$ (A80II).
As expected from the advantages discussed in \S \ref{sec:tf} the scatter is reduced in comparison with the earlier
optical work.
Consideration of more distant galaxies revised this estimate up to $H_0 = 95$ (A80III), with the previous low value
explained by a motion of the local group relative to Virgo, in agreement with the recently discovered
CMB dipole anisotropy \cite{CMB_dipole}.

Further investigation of corrections and scatter in the IR TF, with the inclusion of more clusters, has revised this
further to $H_0 = 82$ km s$^{-1}$ Mpc$^{-1}$ \cite{AM83}, and confirmed the agreement between local peculiar
motions and the CMB dipole \cite{ABMHSC86}.
Another important study, contemporary to Aaronson \etal\ but working in the optical, was a performed by 
Bottinelli, Gouguenheim, Paturel \& de Vaucouleurs.  In particular they investigated the properties of HI
line widths, and demonstrated the importance of correcting for turbulent motions \cite{BGPdV83}. 

The advent of CCD photometry in the late 1980s improved the accuracy of galaxy magnitudes, and there is now
a wealth of data in the $B$, $R$, and in particular, $I$-band.
Before this revolution magnitudes were measured photographically or by single photoelectric detectors through an aperture.
However, the IR is not accessible with CCD cameras.  This is a reason for the popularity of the $I$-band,
the longest wavelength band available to CCDs.
The application of CCD technology to the TF was pioneered by Bothun \& Mould \cite*{BM87} who demonstrated
a reduction in the TF scatter with CCD $I$-band surface photometry.  They also examined the possibility
of a bulk motion out to $3000$ km s$^{-1}$, perpendicular to the supergalactic plane, though finding no evidence
for a flow $> 350$ km s$^{-1}$.
Work by Pierce \& Tully \cite*{PT88} examined the TF parameters in differing environments, concluding that
environmental effects were small.
They also found evidence for substructure in Virgo, and assuming a 300 km s$^{-1}$ flow of the local group
towards Virgo, estimated $H_0 = 85$ km s$^{-1}$ Mpc$^{-1}$.
Both these studies found low scatter in the $I$-band relation, down to $~0.2$ mag, although conclude that the scatter
is still significantly larger than expected from measurement errors.

Around this time it was noticed that the galaxy velocities measured by the TF and other methods were in
disagreement with a uniform Hubble flow, even in the CMB rest frame.  The origins of these peculiar
velocities were suggested to be infall into the surrounding superclusters, and in particular a structure behind
Hydra-Centaurus \cite{GA}, dubbed the Great Attractor.

Several large studies followed. These include an `all sky' survey by Mould and collaborators, finding a reasonably
quiescent Hubble flow in the North \cite{M93} from the $H$-band TF, and some evidence for an outflow in the South
\cite{M91} working in the optical.  Comparing their Northern study with predicted peculiar velocities based on the
IRAS redshift survey, they found general agreement, except in the vicinity of dense structures where the measured
velocities were significantly greater than the IRAS predictions.  An attempt to distinguish between a Great Attractor
model and bulk flow was inconclusive.
Han \cite*{Han92_survey} carried out a large $I$-band CCD survey of 16 nearby clusters and, with Mould \cite[(HM)]{HM92},
5 clusters in the Perseus-Pisces supercluster.  They found the data to fit a model with infall into
both Perseus-Pisces and Hydra-Centaurus at least as well as the bulk flow model,
but again were unable to distinguish between these alternatives.
In the same year Mathewson, Ford \& Buchhorn \cite[(MFB)]{MFB92} performed a large southern $I$-band cluster survey, 
using a combination of HI radio line widths and velocity widths derived from H$\alpha$ optical long-slit spectroscopy.
This latter technique was developed further by Schommer \etal\ \cite*{SBWM93} using an imaging spectrometer
and fitting the resulting 2-dimensional velocity fields.  This study also confirmed large peculiar velocities
in the direction of Hydra-Centaurus.
Meanwhile, Pierce \& Tully \cite*{PT92} investigated the absolute calibration of the TF in detail,
finding some systematic differences between local calibrator galaxies and those in distant clusters,
though not in the field.  This suggested evolutionary problems with absolute calibration.

Giovanelli \cite*{G95} examined the dependence of various galaxy properties with luminosity,
and in particular found a significant dependence for internal extinction.
Willick \cite*{Willick94} has also done much to improve understanding of biases in the TF, with a detailed
examination of selection effects and methods to correct for them.

At this point two main studies attempted to combine the large amount of data available from different sources
into one homogeneous sample.  The first such example is the Mark III catalogue by Willick and collaborators.
This combined TF data from several of the sources listed above, in the $r$, $I$ and $H$-band with both HI and
H$\alpha$ velocity widths, and adjusted them in an attempt to make the data homogeneous.
The calibration of cluster (Mk3-I) and field (Mk3-II) galaxies
are treated separately, and then combined, together with $D_n - \sigma$ data (Mk3-III).
This catalogue has been analysed using the POTENT method \cite{MarkIII-V} of Dekel, Bertschinger \&
Faber \cite*{POTENT90},
finding a bulk flow (with respect to the CMB) within 50 $h^{-1}$ Mpc of over 300 km s$^{-1}$.
The Mark III TF data has also been compared with the peculiar velocities
predicted by IRAS, using a maximum likelihood method, termed VELMOD \cite{MaxL-I,MaxL-II}, in an attempt
to find cosmological constraints from the relation between the peculiar velocity and density fields.

The second large study is that of \Gio, with their particular combination of data
from various sources and their own observations.  The data and analysis is divided into a 
field sample (SFI) and a cluster sample (SCI) \cite[(G97I), Table 2]{Gio97_data}.
The SCI sample is used in this project, and is discussed in more detail in \S \ref{sec:sample}.
A thorough TF analysis of the SCI sample \cite[(G97II)]{Gio97_analysis}, 
and subsequent examination the derived cluster motions out to $9200$ km s$^{-1}$ \cite[(G98)]{G98cluster},
found a small bulk flow of approximately $250$ km s$^{-1}$ within $6000$ km s$^{-1}$, but none 
when considering the sample $> 3000$ km s$^{-1}$.  That is, a local convergence of the velocity field.
This is contrary to the conclusions of the Mark III catalogue \cite{MarkIII-V}.
The G97I/II analysis determines an absolute calibration by assuming the distant velocity field to have an
average zero peculiar velocity.  This allows detection of all bulk flows, except for a monopole component.
Hence, this method cannot determine $H_0$, nor it's isotropic variation with distance.
$H_0$ is found by employing calibrator galaxies with distances determined by other means.
In this case, and most others, Cepheid distances are used to give an absolute calibration.
For the SCI sample this gives $H_0 = 69$ km s$^{-1}$ Mpc$^{-1}$ \cite[(G97III)]{G97_H0}.

Tully, Pierce and collaborators have also conducted an analysis similar to that of G97I/II on their
own combined sample, primarily in the $B$, $R$ and $I$-bands, but also with limited $K'$-band photometry
\cite{TP2000}.  They find $H_0 = 77$ km s$^{-1}$ Mpc$^{-1}$, in disagreement with the G97III
analysis, despite a large overlap of $I$-band data.  Attempts to adjust for the differing
analysis methods still leaves a $\sim$$5$\% discrepancy.

The survey of Dale \etal\ \cite*{Dale99_V} expands on the
Giovanelli \etal\ work with $I$-band photometry and optical rotation curve measurements of 52 Abell clusters
out to 200 $h^{-1}$ Mpc, in order to accurately determine the bulk flow on a 100 $h^{-1}$ Mpc scale.
This survey provides a homogeneous data set from observations made purely for this purpose, 
improving confidence in the results of this analysis in comparison with those combining data
from varied sources. Dale \etal\ \cite*{Dale99_III} finds no bulk flow greater than 200 km s$^{-1}$.

The HST Key Project team have conducted their own analysis \cite{Sakai_etal}, using a combination of their own 
$B$, $V$, $R$ and $I$-band observations \cite{Macri_etal} with other surveys, including primarily the SCI sample.
They also include $H$-band photometry by Aaronsen \etal\  They calibrate using Cepheid distances
to 21 galaxies, and find $H_0 = 71$ km s$^{-1}$ Mpc$^{-1}$.

It is thought that many of the discrepancies between bulk flows and peculiar velocities measured by
different groups are caused by the sample combination procedures. This has been demonstrated for the Mark III
catalogue by Shellflow \cite{Shellflow}, a recent analysis by the creators of the Mark III.
Shellflow is a uniform $I$-band, H$\alpha$, all-sky TF survey on a narrow redshift band 
centred at $\sim$$6000$ km s$^{-1}$.
No significant motion of this shell is found with respect to the CMB frame.
Hence agreement is starting to be reached, with convergence of the velocity field at around 50 $h^{-1}$ Mpc
measured by many TF studies.

Electronic cameras capable of imaging in the IR have taken longer to develop than their optical counterparts,
and simple aperture photometry has been the only option until recently.  
IR work has also been limited due to the problems of strong and variable atmospheric
emission at these wavelengths.  Obtaining consistent photometry thus requires a substantial effort.
Most work on the IR TF has been done in the $H$-band, and in particular using the $H_{-0.5}$ 
aperture magnitude, introduced by Aaronson, Mould \& Huchra \cite*{AMH79}.
Because of the limitations discussed this magnitude is defined in a hybrid manner,
as the integrated flux within an aperture of diameter $10^{-0.5} D_{25}$, where $D_{25}$ is the diameter of
the \emph{optical} $25$ mag arcsec$^{-2}$ isophote.
This is clearly not ideal.  Indeed, Sakai \etal\ \cite*{Sakai_etal} raise doubts as to the consistency of 
this magnitude, finding an offset between nearby and distant galaxies when compared with optical bands.
However, further investigation by Watanabe \etal\ \cite*{WYIIY01} has found this effect remains
even if $H$-band surface photometry is used.

Electronic imagers for IR wavelengths became available in the mid-1990s.  Although initially limited to small
fields, their size has increased over the following years.
It is therefore only recently that the more self-consistent
and sophisticated magnitudes available with imaging have been applied in the IR \cite{RSTW00,WYIIY01}.
The most significant use of this technology is the 2MASS survey \cite{2MASS_Skrutskie}, discussed in \S \ref{sec:2mass}.
An initial study into using 2MASS data with the TF relation has been performed by Bouch{\' e} \& Schneider
\cite*{BS2000}, discussed later in this report.  In addition 2MASS data has been applied to the related
Fundamental Plane method \cite{Hogg_Thesis}.  These two studies form some of the motivation for this project.

\subsection{Outline of this study}
This study aims to evaluate the utility of applying the TF relation using 2MASS data.
2MASS provides near-infrared photometry in the $J$, $H$ and $K_s$-bands.  The project and
it's data products are described in detail in \S \ref{sec:2mass}.
In order to achieve this study's aim, a cluster TF analysis is performed, using the previously
collated SCI sample data set (G97I) for the velocity widths.  The sample is discussed in \S \ref{sec:sample},
its velocity widths in \S \ref{sec:velwidths}, and its inclinations in \S \ref{sec:inclinations}.
SCI also provides $I$-band photometry, which is processed along with that of 2MASS for comparison.
Details of the photometry and corrections applied to it are given in \S \ref{sec:photometry}.
The method used to fit the TF relation requires care.  This is discussed, and my chosen process described,
in \S \ref{sec:applying}.
\Gio\ (G97II) conduct their own analysis of the SCI data set, and the process and results of their study,
are compared with my own.
Cluster distances and peculiar velocities are presented in \S \ref{sec:results}.  The $I$-band results are contrasted
with those from the 2MASS data, and these are all compared to the results of \Gio\ (G98).
Finally \S \ref{sec:conclusions} presents this study's conclusions as to the utility of the 2MASS TF
and prospects for future work.

\section{2MASS} \label{sec:2mass}
\subsection{The 2-Micron All Sky Survey}
The Two Micron All Sky Survey (\2mass) is a ground-based, all-sky survey in the near-infrared bands 
$J$ ($1.25\ \mu$m), $H$ ($1.65\ \mu$m) and $K_s$ ($2.17\ \mu$m).
It is described in detail in documentation available on the world wide web%
\footnote{from http://www.ipac.caltech.edu/2mass/}.
It aims to provide highly reliable and uniform all-sky NIR photometry.
To meet this aim the survey was performed by two dedicated 1.3 m telescopes, 
one each for northern and southern declinations, and the entire observation and data reduction 
process has been conducted in a consistent manner, with a rigorous quality control system.
The entire data acquisition part of the survey has been completed as of the end of October 2000,
and the data-reduction process is currently underway.   

The products of the \2mass\ survey are; a multi-band digital atlas of the sky, 
a point source catalogue (PSC) with positions and fluxes for $\sim$$300$ million stars and 
other unresolved objects,
and an extended source catalogue (XSC) containing positions and magnitudes for more 
than $1,000,000$ galaxies and other nebulae. 

There have been two incremental releases to date, in Spring 1999 and
Winter 2000, in order that the data be applied as soon as possible,
and for problems to be identified and corrected before the final release.
These incremental releases only cover a part of the total sky.
The second incremental release of the XSC is used in this project, covering 47\% of the sky.
The main documentation for this latest 2MASS release is the 
`Explanatory Supplement to the 2MASS Second Incremental Data Release'
\cite{2MASS_exp_sup}
The final, complete release is anticipated in late 2002.

\subsection{The Extended Source Catalog} \label{sec:XSC}
The contents of the 2MASS XSC, and the algorithms used to create them,
are described in Jarrett \etal\ \cite*{2MASS_XSC}.
The extended sources --- mostly galaxies --- are identified in the survey images, extracted, 
and each analysed to provide a large number of data items.  This is done by an automatic data pipeline
known as GALWORKS.

The data items produced include many different magnitude measurements.
All magnitudes are labelled `$<$band$>$\_m\_$<$type$>$',
where $<$band$>$ is one of `j', `h', or `k', and $<$type$>$ is specified in the text below.
The simplest measures are the integrated flux within fixed circular apertures with a range of radii.  
The $<$type$>$ label for these measurements is the radius of the aperture in arcsec.

The rest of the magnitudes are measured within adaptive apertures.
Firstly, there are `individual' measurements.  
In each band a circle and an ellipse are fit to both the $20$ and $21$ mag arcsec$^{-1}$ isophotes.  The flux integrated over each
of these apertures gives a magnitude.  The $<$type$>$ appearing in the label is constructed from an `i', for individual, 
the number corresponding to the isophote, and either a `c', for circular, or an `e', for elliptical.  
For example, the $H$-band magnitude measured within an elliptical aperture fit to the $20$ mag arcsec$^{-1}$
($H$-band) isophote is labelled `h\_m\_i20e'.
There are also individual Kron magnitudes \cite{Kron80}, both circular and elliptical, just denoted by $<$type$>$ labels 
`c' and `e' respectively.  These measure the flux within an aperture controlled by the first moment of the light
distribution, and are thought to be a more robust measure of the total magnitude.

Secondly, and more usefully, there are `fiducial' measurements.  These are done in the same way as the `individual' 
measurements, except that the aperture is determined in one band and then this same aperture used to 
measure the magnitude in the other bands.  This is done with an aperture fit to both the $K_s$-band 
$20$ mag arcsec$^{-1}$ isophote and the $J$-band $21$ mag arcsec$^{-1}$ isophote.
This gives a comparable set of magnitudes for the three bands, useful for reliable determination of colours.  
The `fiducial' measurements are labelled with a $<$type$>$ of the form `k20f' or `j21f' followed by an `e', for elliptical,
or a `c' for circular.  For example, the $K_s$-band magnitude measured within the fiducial elliptical aperture fit 
to the $21$ mag arcsec$^{-1}$ $J$-band isophote is labelled `k\_m\_j21fe'.
Again, there are also fiducial Kron magnitudes, with $<$type$>$ `fc' or `fe'.

The centre of all apertures is the peak intensity pixel for the source.
The ellipse parameters for elliptical apertures are determined by fitting an ellipse to the $3\sigma$
isophote in the band considered.  This is the isophote corresponding to a surface brightness of 3 times
the background noise, in practice fixed to typical levels of $20.09$ mag arcsec$^{-1}$ in $J$, 
$19.34$ mag arcsec$^{-1}$ in $H$, and $18.55$ mag arcsec$^{-1}$ in $K_s$.  These ellipses are then scaled to
best fit the required isophote, or to the calculated Kron radius.

The 2MASS XSC has guaranteed specifications in unconfused regions of the sky (mostly away from the Galactic plane)%
\footnote{http://www.ipac.caltech.edu/2mass/releases/second/doc/requirements.html}.
The completeness level is the fraction of all galaxies present, brighter than the magnitude limit,
that are included in the catalogue.  The reliability gives the fraction of galaxies included in the catalogue
that are truly extended sources rather than, for example, mis-identified triple stars.
The extended source magnitude limits (signal-to-noise ratio, $S/N > 10$) are $15.0$, $14.3$, and $13.5$ mag 
respectively for the $J$, $H$, and $K_s$-bands.  Brighter than these magnitudes and $> 30^\circ$ from the Galactic
plane the XSC has better than 90\% completeness and 99\% reliability.  Below this Galactic latitude a 
completeness level can not be assured, although the reliability remains 99\% down to $|b| > 20^\circ$
and 80\% between $20^\circ > |b| > 10^\circ$ but with no guarantee for $|b| < 10^\circ$. 

\section{The sample} \label{sec:sample}
The initial source of galaxies for this study is the SCI sample of \Gio\ (G97I).
This is a collection of TF data for 782 spiral galaxies, in the fields of 24 clusters or groups.
They comprise $I$-band photometry and a combination of
radio and optical velocity widths.  Discussions of these data are presented in 
\S\S \ref{sec:photometry} \& \ref{sec:velwidths} respectively.
Here we consider the sample selection.  This is important in order to 
correctly assess the selection biases which may affect the analysis.

Galaxies in clusters are required for this analysis, for the reasons discussed in \S \ref{sec:applying}.
The clusters in the SCI sample were primarily chosen as those for which there was enough TF data available.
That is, those with $I$-band photometry and velocity width measurements for a reasonable number of galaxies.
A secondary consideration was that the clusters should be well distributed around the sky, in order for their
average motions to represent the true bulk motion of the region.
This has been moderately well achieved.

The galaxy sample for each cluster is again primarily defined by the availability of data, although as much is
\Gios\ own the sampling should be more understood, however it is unclear how they chose which galaxies to observe.
In their bias correction procedure they fit a completeness histogram w.r.t. absolute magnitude with a smooth
step function (G97II: Figure 10), so it appears they believe their sample to be approximately magnitude limited.
However, given the `gaps' obvious in the completeness histograms of most of the nearest,
and hence deepest sampled, clusters, it seems that either the selection function under-represents galaxies
with intermediate magnitudes in each cluster, or that this lack of intermediate luminosity galaxies
is a real property of the luminosity function that is not accounted for.  We will leave this problem for now.

The SCI sample comprises galaxies in the \emph{fields} of clusters and groups, but their actual physical
membership of the concentration has been assessed in G97I.  They establish four possibilities for each galaxy;
that it be a true member of the concentration, a peripheral member, in the foreground, or in the background.
The galaxies in each cluster are then examined in sky projection and redshift to determine the category to
which they should be assigned.  198 of the SCI galaxies are rejected as fore or background.
374 are classified as true members, forming what they term the \in\ sample.
The remaining 210 galaxies are found to be close to the concentration's mean redshift, but
spatially removed from it's centre and therefore membership cannot assigned with certainty.  These peripheral
galaxies, in combination with the \in\ sample, form the \inp\ sample.
Some clusters with more complex sub-structure are assigned correspondingly more detailed membership codes.
Cancer has been recognised to consist of a number of clumps, dubbed A, B, C, D \cite{BGBH83}.
As with G97I, this study only considers clump A when referring to Cancer.
Centaurus has been found to comprise two concentrations, superimposed on the sky but with approximate
redshifts of $3000$ and $4500$ km s$^{-1}$.  Only members of the $3000$ km s$^{-1}$ structure are assigned
to the \inp\ sample and hence considered in this study.

The pairs of clusters A2197/A2199 and A2634/A2666 have confused membership issues, and are treated
in special ways by G97I/II.  To avoid complicating the analysis process of this study I have chosen
to remove them from the sample considered.

The SCI \inp\ sample with the removals specified is to be combined with 2MASS data for the corresponding
galaxies.  A number of galaxies in SCI are not given a previous catalogue reference. Also the degree to which
2MASS objects are identified with catalogued galaxies is unsure.
The 2MASS objects corresponding to SCI galaxies are therefore identified by position,
rather than catalogue reference.
The positions given in G97I are, however, not particularly accurate. They are therefore improved,
in order to ensure matches with the 2MASS catalogue, which has high astrometric accuracy \cite{2MASS_exp_sup},
that are as reliable as possible.  This was done by obtaining the most accurate positions in NED%
\footnote{the NASA/IPAC Extragalactic Database, see the Acknowledgements section.}
for all the SCI galaxies with catalogue references.
Comparison of the positions thus obtained with those of G97I demonstrates a number of disagreements
greater than an arcminute, although most are under $2$ arcminutes.  All galaxies with differences $>90''$ 
have been examined.  The most serious discrepancies appear to be misidentifications and these galaxies are 
removed from the sample considered.
Inspection of a selection of galaxies with disagreements $>60''$ suggests these
galaxies are all correctly identified (by comparison with properties from NED), but the positions are inaccurate.
These galaxies are therefore allowed to remain in this study's sample.
A list of galaxies inspected, and all those removed due to anomalous positions, appears in 
table \ref{tab:anompos}.

\begin{table}[tb]
\footnotesize
\centering
\hrulefill
\begin{description}
\item [$754$'': ESO 444-G001] Appears to be a mis-identification: {\bf removed}.
\item [$326$'': CGCG 098-001] Position given is that of a different nearby galaxy.
        Treat as mis-id: {\bf removed}.
\item [$159$'': IC 2288] Position incorrect (nearby star) - noted by NED.  Assume id correct.
\item [$97$'': UGC 06911] Nothing at quoted position. Properties similar to NED, assume id correct.
\item [$89$'': UGC 00485] Nothing at quoted position. Properties similar to NED, assume id correct.
\item [$71$'': NGC 1350] Nothing at quoted position. Properties similar to NED, assume id correct.
\item [$61$'': UGC 00809] Nothing at quoted position. Properties similar to NED, assume id correct.
\end{description}
\hrulefill
\caption{\label{tab:anompos}
        Explanations for anomalous \Gio\ -- \2mass\ separations (all of those $>90$'' and some $>60$'').
        All in this table appear to be due to inaccuracies, or mistakes, in the \Gio\ positions, 
        but with identifications being correct.  All galaxies in the sample are therefore assumed to be 
        correctly identified, with the exception of those flagged as removed in this table.}
\end{table}

The 2MASS XSC was searched%
\footnote{using the IRSA Gator Catalog Query tool accessible from http://irsa.ipac.caltech.edu/}
for galaxies within 15 arcsecs of the best positions available (NED or G97I).
However, as stated in \S \ref{sec:2mass}, currently only 47\% of the sky is available, as the
second incremental release.  This coverage is in thin strips distributed in a non-trivial manner across the sky.
The SCI clusters were affected to different levels, but all had some galaxies that were unavailable from 2MASS.
Unfortunately, a number of clusters are depleted to three or fewer galaxies, and therefore lost from the sample.

Although lack of 2MASS coverage is the dominant factor in removing galaxies from the sample, some may be lost
due to the rising 2MASS incompleteness near and beyond the magnitude limit.  The SCI galaxies were compared with a
mask of the 2MASS second incremental release coverage%
\footnote{kindly performed by Robert Hosford, using a mask developed as part of his MSci project \cite{HosfordMSci}.}
in order to assess the proportion of galaxies not matched due to incompleteness 
or being fainter than the 2MASS magnitude limit.  Of the 782 original SCI galaxies, 70 were supposedly within the
2MASS coverage mask but not matched with a 2MASS source.  The average $I$-band magnitude of these is $13.2$ mag,
not very much fainter than the entire SCI average of $12.7$ mag.  The galaxies are typically $\sim$$0.5$, $1$ 
and $1.5$ mag brighter than $I$ in $J$, $H$ and $K_s$ respectively, so the unmatched galaxies are also generally much 
brighter than the 2MASS completeness limits.
There may be a problem with the coverage mask resolution, suggesting the true number of unmatched 
galaxies is much lower.
So, when combining the SCI sample with 2MASS data, the only significant factor reducing the size of the sample
appears to be the limited 2MASS coverage.

\begin{figure}[tb]
\centering
\includegraphics[width=275pt]{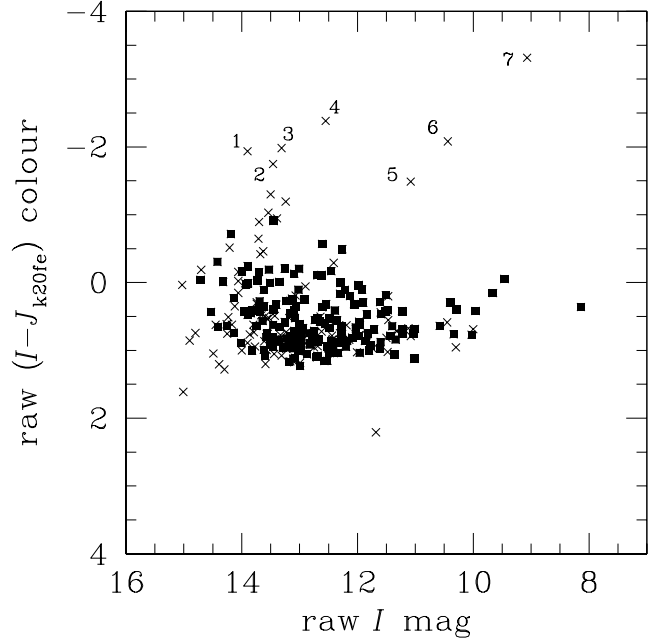}
\caption{\label{fig:colourcomparisons}
Uncorrected $I_{SCI} - J_{k20fe}$ colour versus magnitude, for all of the SCI galaxies with \2mass\ magnitudes
(crosses) and those in the final sample (squares).
The $I - H, K$ plots are very similar.
The labelled outlying points are removed from the sample for the reasons given in table \ref{tab:colourremoved}.}
\end{figure}

The photometry is considered in depth in \S\S \ref{sec:photometry} \& \ref{sec:photocorrections}, 
but at this stage the G97I and 2MASS photometry are compared to identify problems.  
The $I-J$ colour is plotted, using the raw magnitudes, in 
figure \ref{fig:colourcomparisons}.  There is a large scatter, due to intrinsic galaxy colour variations
as well as differences in the way the magnitudes are measured, which increases with magnitude.
However, there are a number of outlying points.  These are examined and many excluded from the sample for the 
reasons given in table \ref{tab:colourremoved}.
2MASS provides flags identifying problem measurements.  These are not used to eliminate points, although this
would probably be useful in future studies with more statistics.

\begin{table}[tb]
\footnotesize
\centering
\hrulefill
\begin{description}
\item [1. NGC 1350] Lies on the edge of a \2mass\ scan area $\Rightarrow$ some flux missing in \2mass measurements.
\item [2. NGC 1744] Large, loose, faint arms $\Rightarrow$ probably hard to isophotally fit with \2mass's 
        low signal-to-noise.
\item [3. NGC 1179] Noisy \2mass\ images.
\item [4. ESO 422-G005] Very noisy \2mass\ images.
\item [5. IC 1962] Very noisy \2mass\ images.
\item [6. ESO 357-G012] Very low signal-to-noise \2mass\ images.
\item [7. ESO 289-G026] Loose, asymmetric arms, strongly-barred $\Rightarrow$ probably inaccurate 2MASS
isophotal fit.
\end{description}
\hrulefill
\caption{\label{tab:colourremoved}
List of galaxies removed from the data set due to poor photometry. The same galaxies are excluded
by both k20fe and kron circular photometry.
The numbering refers to the labels in figure \ref{fig:colourcomparisons}.}
\end{table}

Finally, three galaxies identified as severe outliers in TF plots compiled later are removed from the sample.
G97I includes a long list identifying potential problems with certain galaxies in the SCI sample.  These three
galaxies (UGC 1416, UGC 6693, UGC 4264) are all stated as not used in the G97I fits due to different problems.
Other galaxies excluded by the analysis of G97I are not excluded in this study other than by
the reasons given above.

The final sample used in this study contains 153 galaxies distributed in 11 clusters, as given in table
\ref{tab:clusternumbers}, along with the numbers removed from the initial SCI sample.

\begin{table}[tb]
\footnotesize
\centering
\begin{tabular}{l|r|r|r|r|r|r}
        \hline
        \textbf{Cluster} & {\boldmath $N_{\text{\inp}}$} & {\boldmath $N_{\text{\inp}\ \cup\ \text{2MASS}}$} 
                         & {\boldmath $-N_{\text{mis-id}}$} & {\boldmath $-N_{\text{colour}}$} 
                         & {\boldmath $-N_{\text{?}}$} & {\boldmath $N_{\text{final}}$}\\
        \hline
        \textbf{N 383 Group (Pisces)}   & $21$  & $15$  & $0$   & $0$   & $0$   & $15$ \\
        \textbf{N 507 Group}            & $14$  & $7$   & $0$   & $0$   & $0$   & $7$  \\
        \textbf{A 262}                  & $31$  & $21$  & $0$   & $0$   & $1$   & $20$ \\
        A 400                           & $25$  & $0$   & $0$   & $0$   & $0$   & $0$ \\
        \textbf{Eridanus}               & $34$  & $21$  & $0$   & $2$   & $0$   & $19$ \\
        \textbf{Fornax (S 0373)}        & $39$  & $13$  & $0$   & $4$   & $0$   & $9$  \\
        \textbf{Cancer}                 & $26$  & $19$  & $0$   & $0$   & $0$   & $19$ \\
        Antlia                          & $27$  & $1$   & $0$   & $0$   & $0$   & $1$ \\
        Hydra (A 1060)                  & $25$  & $3$   & $0$   & $0$   & $0$   & $3$ \\
        N 3557 Group                    & $11$  & $0$   & $0$   & $0$   & $0$   & $0$ \\
        \textbf{A 1367}                 & $35$  & $23$  & $1$   & $0$   & $1$   & $21$ \\
        \textbf{Ursa Major}             & $30$  & $10$  & $0$   & $0$   & $0$   & $10$ \\
        Cen30 (A 3526)                  & $38$  & $0$   & $0$   & $0$   & $0$   & $0$  \\
        \textbf{Coma (A 1656)}          & $41$  & $19$  & $0$   & $0$   & $0$   & $19$ \\
        \textbf{ESO 508}                & $17$  & $11$  & $0$   & $0$   & $0$   & $11$ \\
        A 3574                          & $20$  & $5$   & $1$   & $0$   & $1$   & $3$ \\
        A 2197/9                        & $25$  & $3$   & $0$   & $0$   & $all$ & $0$  \\
        Pavo II (S 0805)                & $18$  & $0$   & $0$   & $0$   & $0$   & $0$ \\
        Pavo                            & $10$  & $0$   & $0$   & $0$   & $0$   & $0$ \\
        \textbf{MDL 59}                 & $23$  & $5$   & $0$   & $1$   & $0$   & $4$  \\
        Pegasus                         & $17$  & $2$   & $0$   & $0$   & $0$   & $2$ \\
        A 2634/66                       & $28$  & $3$   & $0$   & $0$   & $all$ & $0$  \\
        \hline
\end{tabular}
\caption{\label{tab:clusternumbers}
Numbers of galaxies in the sample for each cluster.
$N_{\text{\inp}}$ is the number of galaxies in the \inp\ sample,
$N_{\text{\inp}\ \cup\ \text{2MASS}}$ is the number left when combined with 2MASS,
$N_{\text{mis-id}}$ is the number removed because of mis-identification,
$N_{\text{colour}}$ is the number removed because of photometry problems,
$N_{\text{?}}$ is the number removed because of other reasons (see text), and
$N_{\text{final}}$ is the total number remaining in the final sample.
Those clusters remaining in the sample ($N_{\text{final}} > 3$) are in bold type.}
\end{table}

\section{Velocity widths} \label{sec:velwidths}
The velocity widths are entirely from the SCI sample (G97I), and originally collated from \Gios\ own
observations and a number of other sources.
They are either the values measured at the 50\% level of the radio profile horns, 
or those obtained from optical, single-slit photometry, corrected to make them equivalent to the radio
measurements.  Detailed discussion of this process is beyond the scope of this report.
The corrections applied by \Gio\ are are assumed to be sufficiently accurate, 
and the fully corrected values from the SCI, $\Wcor$, are used directly in this project.
These corrections are briefly summarised here.

The HI radio measurements are corrected for instrumental broadening and smoothing, which may have been
applied to improve the S/N, by subtracting an instrument (and smoothing method) dependent term,
$\Delta_s$, from the observed line width, $W_{\text{obs,21}}$.  This is then multiplied by a factor
$(1+z)^{-1}$ to account for broadening by relativistic effects.
Next a term describing broadening due to turbulent motions in the disk is subtracted in quadrature,
and finally the disk is deprojected by a factor $\sin^{-1}{i}$.  The inclination used in this correction
is discussed in \S \ref{sec:inclinations}.
The optical H$\alpha$ measurements require a similar treatment, except there is no correction for
turbulence, and the $\Delta_s$ correction becomes $\Delta_{sh}$, an offset to make the measurement
equivalent to a corrected HI measurement.

For more details see \Gio\ \cite*[\S5]{Gio97_data} and references therein.
Unfortunately, \Gio\ does not quote errors on the raw measurements, but does so for the final corrected 
velocity widths, and, again, these are assumed to have been accurately assessed.
These errors are typically $< 10$\% but vary between $\sim$$1 - 20$\%.

\section{Inclinations} \label{sec:inclinations}
Inclinations are determined by fitting an ellipse to an isophote of the galaxy image.
The ellipticity is then $e = 1 - b/a$, where $a$ and $b$ are the major and minor
axes of the ellipse.
For there own data G97I determines the mean ellipticity, $\bar e$ for the range of radii $r = a/2$\ for which 
the disk surface brightness (SB) is approximately exponential.  This is then corrected for the smearing effects
of seeing to give the corrected ellipticity, $e$. (See \S4 of G97I for more detail.)

If the disk had no height, then the inclination angle would just be ${\cos{i}} = 1-e$.
However, the disks do have an intrinsic axial ratio, $q_0$, and may be modelled as highly oblate spheroids.
G97I adopts $q_0 = 0.13$ for type Sbc and Sc galaxies and $q_0 = 0.2$ for all others.  However, many other
sources adopt $q_0 = 0.2$ for \emph{all} types. This presents a problem when converting types Sbc and Sc to
the $q_0 = 0.13$ system as the maximum inclination is then $81^\circ$.  If the inclination exceeds this,
G97I recomputes the inclination for data with available axial ratio measurements, but leaves unchanged
data without axial ratios.  However the inclination dependent corrections (to velocity widths, 
see \S \ref{sec:velwidths}) are small near $i = 90^\circ$ so this discrepancy is not important.
No seeing corrections are applied in G97I to data from other sources.

The ellipticity errors quoted by G97I vary from $\sim$$8$\% at $e \sim 0.2$ to $\sim$$2$\% at $e \sim 0.8$,
but display considerable scatter, to a maximum error of $\sim$$15$\%.
Note that, for example, at $e = 0.5 \Rightarrow i = 62^\circ$ an error of 15\% in ellipticity
produces a comparable fractional error in the velocity width inclination correction, and hence
in the velocity width itself.

G97I quotes only calculated inclinations, and not the measured axial ratios.
These are required for the internal extinction correction (\S \ref{sec:corintext}) and are therefore
deduced by reversing the G97I procedure given above.
2MASS also provides axial ratios from its ellipse fitting procedure.  These may be useful, but are unlikely
to be of comparable quality because of the low $S/N$.

\section{Photometry overview} \label{sec:photometry}
The $I$-band magnitudes used in this study are entirely from the SCI data set.  These are originally
from a number of sources, in particular \Gios\ own observations and those of Mathewson, 
Ford \& Buchhorn \cite*[(MFB)]{MFB92} and Han \& Mould \cite*[(HM)]{HM92}.  All are derived from 
$I$-band CCD surface photometry, and are estimates of the total magnitude. 
When a scale length is measured for the surface brightness (SB), 
these total magnitudes are generally found by extrapolating the growth curve of the
isophotal magnitudes to infinity, assuming the SB falls exponentially.
G97I applies this extrapolation beyond an isophote between $23.5$ and $24.0$ mag arcsec$^{-1}$.
G97I finds a significant offset ($\sim$$0.05$ mag) between their own total magnitudes and those of MFB, 
for which both have measurements.  
They fit this offset quadratically and adjust the MFB magnitudes to match their own.
None of the other samples has enough overlap to allow comparison, and the total magnitudes from these
sources are included without adjustment.
In cases where a galaxy has more than one measurement an average is usually taken.  These are detailed
in \S3 of G97I.
These total magnitudes are referred to as the `raw' $I$-band magnitudes in this project.

The $J$, $H$ and $K$-band magnitudes for this project are taken entirely from the 2MASS XSC catalogue.
However there are a number of different types of magnitude available, as described in \S \ref{sec:XSC}.
Burkey \cite*{Burkey01} has examined and compared the 2MASS magnitudes.
The fixed aperture magnitudes have the smallest measurement error, mostly just photon noise. They are,
however, clearly biased as estimators of total magnitude.  For larger galaxies the outer flux is not included,
while measurements of small galaxies include all the galaxy's flux plus additional sky background.
Isophotal magnitudes adapt to the size of the galaxy, thus much reducing this bias.  They do introduce an
additional error, though, that of determining the isophotal radius.
They also do not account for differing SB profiles.  Galaxies with flatter SB profiles
will have more flux outside the isophotal aperture than those with steep SB profiles.

The Kron magnitude \cite{Kron80} attempts to better estimate the total magnitude.  In this system the aperture
is scaled to a radius, $r_{\text{Kron}}$, equal to the first moment of the surface brightness, $\mu$,
so that $r_{\text{Kron}} = <\mu r> / <\mu>$, where $< >$ donates the mean value within some area.
For 2MASS this area is that within a radius of roughly $4$ times the exponential scale function.
This system would appear to be more susceptible to contamination by nearby stellar sources,
but a comparison of this with an extrapolated measurement, by Burkey \cite*{Burkey01}, suggests the Kron
magnitudes are more robust than would be expected.  Despite this, stellar contamination is still the 
largest source of error in the Kron magnitudes.

\begin{figure}[tb]
\centering
\includegraphics[width=200pt]{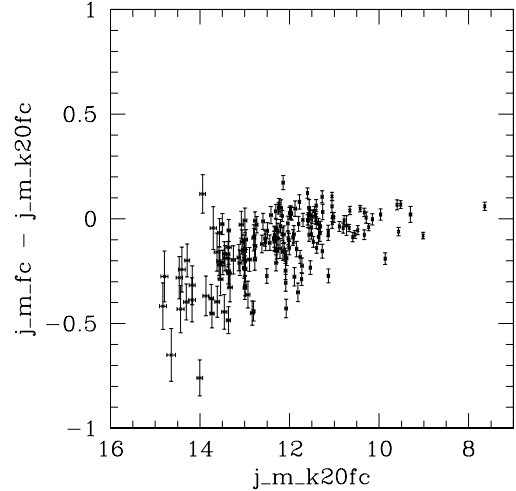}
\caption{\label{fig:kron_v_iso}
Plot of the difference between $J$-band, $K$-fiducial, Kron and isophotal magnitudes, against magnitude.
The scatter, with occasional outliers probably due to contamination errors, increases for fainter
magnitudes.  Also note the tendency for Kron magnitudes to be brighter than the isophotal equivalent at faint
magnitudes.}
\end{figure}

Figure \ref{fig:kron_v_iso} compares the $J$-band circular Kron and `k20fc' magnitudes.  These are similar for
bright galaxies, but below $J \sim 12$ mag the Kron magnitudes become steadily brighter.
This suggests that either the Kron magnitudes oversample or, more likely, that the isophotal
magnitudes undersample, the total magnitudes of faint galaxies.
This would be expected to have an effect on the TF relation, with Kron magnitudes giving a flatter slope.
Both the Kron and isophotal apertures are limited to a minimum size of 7''.  That they display significant
differences at the faintest magnitudes of this sample indicates this minimum has not been reached.

Burkey \cite*{Burkey01} estimates the isophotal and Kron magnitudes underestimate the true total magnitude 
by $10 - 20$\% and $8 - 10$\% respectively.  Note the smaller range of uncertainty for the Kron apertures.
Burkey also tests the repeatability of the magnitude types and finds limiting $K$-band magnitudes
above which the error on repeated measurements exceeds $0.1$ mag. These are $13.5$ mag for fixed aperture,
$13.4$ for isophotal, $13.2$ for Kron and $12.5$ for a simple extrapolated magnitude.

\begin{figure}[tb]
\centering
\subfigure{
        \label{fig:iso_ell_v_cir:cir}
        \includegraphics[width=200pt]{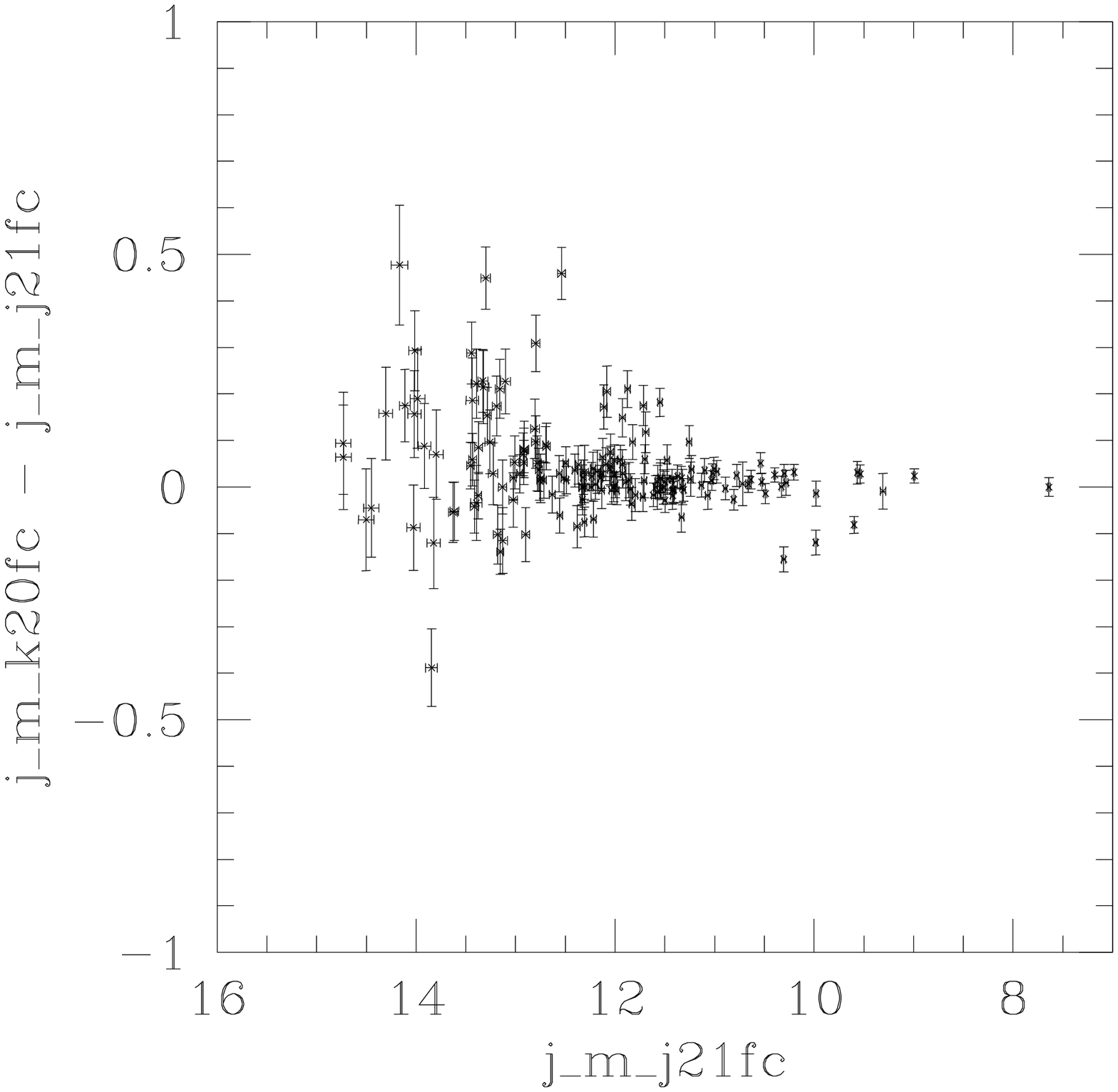}
}
\hfill
\subfigure{
        \label{fig:iso_ell_v_cir:ell}
        \includegraphics[width=200pt]{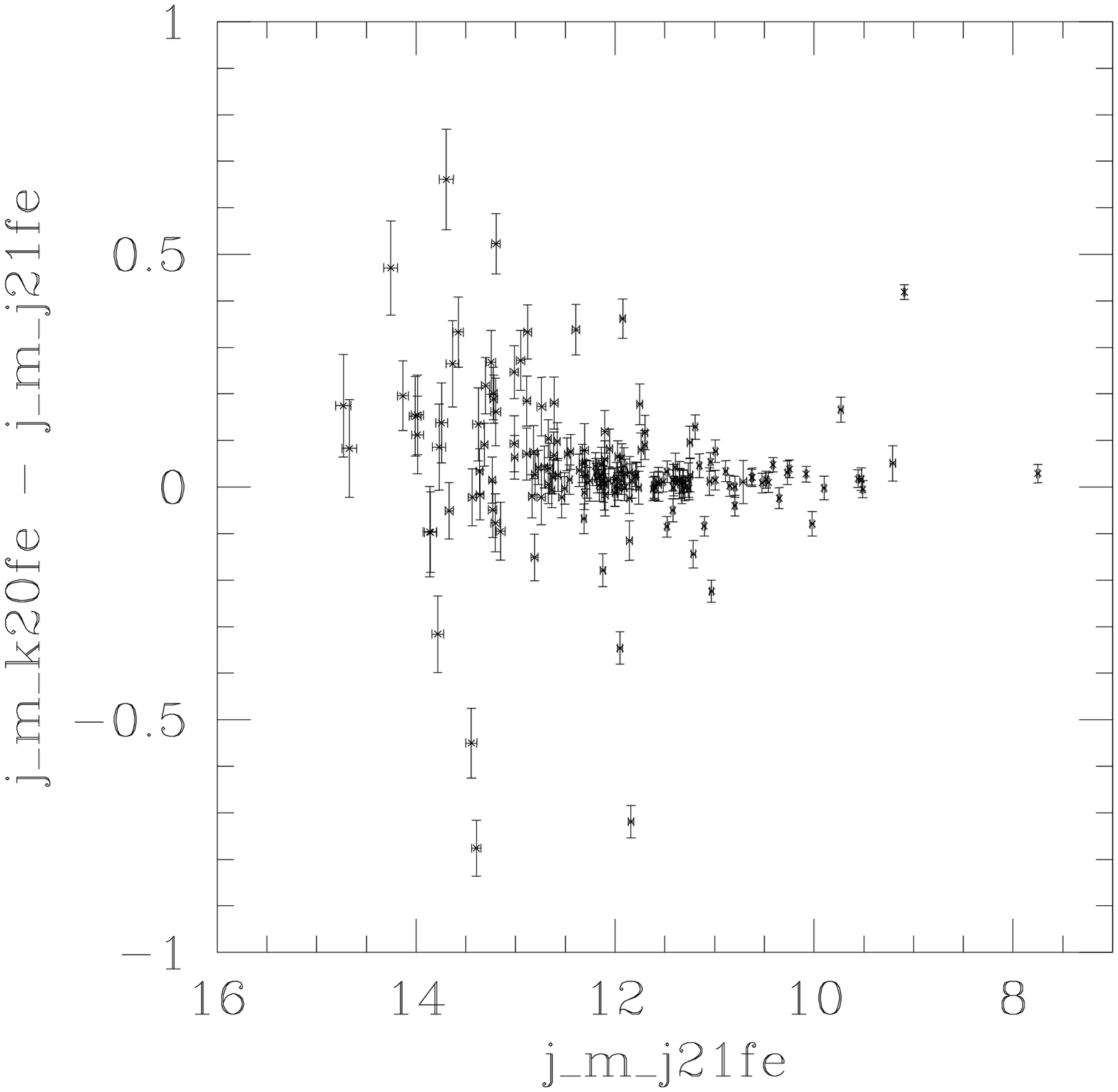}
}
\caption{\label{fig:iso_ell_v_cir}
Plots of the difference between the $J$-band `k20' and `j21' fiducial isophotal magnitudes,
against magnitude.
The left-hand plot uses only circular apertures, while that on the right uses 
only elliptical apertures.  In both plots, points can be seen that are unexpectedly far from the mean difference,
when compared with their error bars.  These outliers are probably caused by contamination of one isophote,
but not the other.  The scatter also increases markedly at larger magnitudes ($J \gtrsim 12.5$ mag), for which the
isophote fitting becomes more uncertain.  Note that the elliptical apertures give more outlying points than the
circular apertures, and in particular that they extend to brighter magnitudes.}
\end{figure}

Elliptical apertures add errors in fitting the angle and axial ratio to those discussed above,
and are also more open to contamination by stellar sources affecting the ellipse parameters.
These errors are hard to evaluate.
Most galaxies appear elliptical on the sky, though, and therefore these apertures reduce inclusion of
background and reduce the possible bias between galaxies with different inclinations.
For small sources in particular the ellipse parameters become highly uncertain.
Figure \ref{fig:iso_ell_v_cir} demonstrates the increase in contamination errors for elliptical
apertures, and the large increase in scatter for both types, towards higher magnitudes, beyond that expected from the
standard error treatment assuming no contamination.

Jarrett \etal\ \cite*{2MASS_XSC} recommends the `k20fe' magnitude as most reliable and robust for galaxy photometry,
but only for brighter galaxies.  For galaxies with $K_s \lesssim 13$ mag they recommend the $7$'' fixed 
circular aperture.  
This is clearly impractical as we require a consistent magnitude measurement for all galaxies. 
Bouch{\' e} \& Schneider \cite*{BS2000} have examined the feasibility of using 2MASS data with the TF relation.
They conclude the $K$-band to be preferable, and find the `k20fe' magnitude type to give the
lowest TF scatter, followed by the $K$-fiducial Kron magnitude.
They only consider galaxies with $K_s \lesssim 13$ mag however.

With little to choose between the different types at the start of this project,
the magnitude primarily worked with in this project is the `k20fe' type.  Other magnitudes are applied in the
later stages for comparison.
Further examination has indicated that the Kron circular aperture magnitude may 
be a promising alternative, for its greater robustness and possibly better sampling of faint galaxies, for which
it was initially designed \cite{Kron80}.
It should also be noted that the use of fiducial magnitudes retains
some of the hybrid character thought to be a disadvantage in the classic $H_{-0.5}$ magnitude measurement, 
discussed in \S \ref{sec:previous}.  In this project the different bands are treated independently, and fiducial
measurements offer no advantage other than that the 2MASS isophote fitting appears to be more reliable in $K$.

In all of this project's analysis the G97I $I$-band data is treated in parallel with the 2MASS $J$, $H$ and $K$
data.  This allows direct comparison of this analysis with that of G97I, G97II and G98.
The raw apparent magnitudes obtained from G97I and 2MASS must be corrected for several effects to improve the 
TF relation and avoid biases.  These are described in the following section.  G97I applies its own
corrections to the $I$-band data, but these are used only for comparison with this project.

\section{Photometry corrections} \label{sec:photocorrections}
\subsection{Cosmological corrections} \label{sec:corcos}
As light travels through the universe the individual photons lose energy to the Hubble expansion.  
This is known as cosmological redshift, as the lower photon energies translate into longer
wavelengths, and hence `reddening'.
Light observed at a particular wavelength will have been emitted at a shorter, distance dependent, 
wavelength.  Sources with the same emission spectrum, but at different distances, will thus in general have 
different observed luminosities at a particular wavelength, dependent upon this emission spectrum.
This effect is accounted for by the $k$-correction. It is complicated, though generally small,
and usually approximated linearly as $kz$.

The $k$-correction in the NIR is one focus of a recent study by Mannucci \etal\ \cite*{Mannucci01}.
They tabulate $k$-corrections for the $J$, $H$ and $K$-bands, out to large $z$ and for different 
morphological types.  This study only requires the most local values, reproduced in table \ref{tab:kcorr}.
There is some variation over morphological type, especially in $J$ and $H$, but to avoid complication this 
study takes and average over the later (Sb \& Sc) types which constitute the majority of the sample. 
The $k$-correction is always very small, being at most $0.004$ mag, and essentially negligible in light
of the other photometric errors present.

\begin{table}[tb]
\centering
\begin{tabular}{l|rrrr}
        \hline
        \textbf{Type} & {\boldmath$I$} &{\boldmath$J$} & {\boldmath$H$} & {\boldmath$K$}\\
        \hline
        Sa      &       0.160   &       0.005   &       -0.036  &       -0.151 \\
        Sb      &       0.160   &       0.017   &       -0.026  &       -0.151 \\
        Sc      &       0.160   &       0.022   &       -0.035  &       -0.155 \\
        \hline
        Used    &       0.160   &       0.020   &       -0.031  &       -0.153 \\
        \hline
\end{tabular}
\caption{\label{tab:kcorr}
Values of $k$ for the cosmological photometry correction from sources.  The bottom line gives the values 
used for all types in this project. $I$-band from G97I, NIR from Mannucci \etal\ (2001) valid for $z < 0.05$.}
\end{table}

\subsection{Galactic extinction} \label{sec:corgalext}
Galactic extinction is the observed dimming of an object due to the absorption of light by dust within the
disk of our Galaxy.
The amount of extinction depends approximately upon the path length the light must travel through the Galactic 
disk before being observed.  However, the dust content of our Galaxy is not smoothly distibuted. 
At medium scales the distribution is `clumpy', and at high resolutions shows a filamentary structure.
More accurate estimates must therefore use a detailed map of the extinction with respect to the
direction out of the galaxy.  At present, the map considered most accurate for this purpose is that by 
Schlegel, Finkbeiner \& Davis \cite*{Schlegel_etal}.  These are obtained from NED, which uses the conversions
of Cardelli \etal\ \cite*{Cardelli_etal} to calculate the extinction in each band.
Note that most previous studies, including G97I, use the older, lower resolution
maps by Burstein \& Heiles \cite*{BH78}.

This correction is applied on a cluster basis for efficiency, and varies from $0.025 - 0.350$ mag in $J$.
In $K$ the extinction is over 5 times smaller, between $0.005 - 0.066$ mag.

\subsection{Internal extinction} \label{sec:corintext}
This is the most complicated photometry correction.
Spiral galaxies contain a variable quantity of dust in their disks.  This scatters and absorbs some of the light as it
travels through the galaxy, causing extinction.  Obviously the further the light must travel through the disk of the galaxy,
the more extinction there will be.  It is therefore dependent upon the inclination at which we observe a galaxy.
Here we define inclination in the same way as \Gio, as the angle between the normal to the disk plane and the line of sight.
Thus an inclination of 0\deg\ corresponds to face-on viewing, and 90\deg\ to edge-on.
Therefore, for larger inclinations the extinction effect is greater, the galaxy appears dimmer, and hence
its total magnitude increases.

The dust content of a galaxy is unevenly distributed in the disk, being dependent upon radius and height above the
disk plane.  An analytic form for extinction \wrt\ inclination or axial ratio is therefore difficult.  The
quantity and distribution being variable between galaxies also suggests an empirical, statistical approach.   
The internal extinction correction used in this project is based upon that presented by Tully \etal\ \cite*{Tully_extinction}.
The extinction, $A_{\lambda}^{\circ}$, is described by a standard empirical function, dependent upon the axial ratio,
$(a/b)$, and a parameter, $\gamma_{\lambda}$.
The parameter, and hence the extinction, are wavelength dependent,
denoted by the subscript $\lambda$.  The superscript ${\circ}$ implies the correction is to the face-on orientation.
\begin{equation}
A_{\lambda}^{\circ} = \gamma_{\lambda} \log{(a/b)}
\end{equation}
The form and value of $\gamma_{\lambda}$ varies between studies.  All studies of internal extinction somehow fit
a form of $\gamma_{\lambda}$ to a set of data.  Han \cite*{Han92} and others use a semi-constant form dependent 
only on morphological type.
Others find a dependency on absolute magnitude.  However, this is usually re-parameterised 
as a function of line-width using the TF correlation \cite{Tully_extinction,Gio97_data}!
The motivation for this re-parameterisation is to remove any distance dependence that could bias
the correction by requiring \emph{a priori} assumed distances.

Tully \etal\ \cite*{Tully_extinction} provide $\gamma_{\lambda}$ as a function of absolute magnitude, 
and re-parameter\-ised as a function of line-width, for the $B$, $R$, $I$ and $K'$ bands.  
However, their line-widths are defined as the width at 20\% of the
horn maxima, whereas those provided by \Gio\ are measured at 50\% of the horn maxima.
In this project the $\gamma_{\lambda}$ as a function of absolute magnitude are used, and the entire process 
iterated over to remove the distance dependence, as described in \S \ref{sec:applying}.
This method is used to avoid introducing additional uncertainty, scatter and bias through both 
a previously calibrated correlation between 
absolute magnitude and line-width, and also through a conversion from 50\% to 20\% line-widths.  
In addition it keeps the axes of the TF relation independent.

The general equation for $\gamma_{\lambda}$ is,
\begin{equation} \label{eqn:gammas}
\gamma_{\lambda} = \gamma^{(1)}_{\lambda} ( \gamma^{(2)}_{\lambda} + M^{k,b,i}_{\lambda} + 5 \log{h_{0}} )
\end{equation}
where $h_0 = H_0 / (100 \text{ km s}^{-1}\text{ Mpc}^{-1}) = 0.75$ is assumed in this project.
Tully \etal\ \cite*{Tully_extinction} gives values of $\gamma^{(1)}_{\lambda}$ and $\gamma^{(2)}_{\lambda}$ for the 
$B$, $R$, $I$ and $K'$ bands.  The correction for $K_s$ is assumed to be the same as that for $K'$.
In addition this study requires corrections for the $J$ and $H$ bands.  These are found by linear interpolation against an
empirical function of $\lambda$, selected by trying several simple functions and choosing the ones for which the points
fit best. For $\gamma^{(1)}_{\lambda}$ this is $\lambda^{-1}$, and for $\gamma^{(2)}_{\lambda}$ the
function used is $\lambda^{-1/2}$.
The $\gamma^{(1,2)}_{\lambda}$ are listed for reference in table \ref{tab:gamma}. 

\begin{table}[tb]
\centering
\begin{tabular}{c|r|r|r}
        \hline
        \textbf{Band} & {\boldmath$\bar\lambda$}\textbf{, \AA} & {\boldmath$\gamma^{(1)}_{\lambda}$} & {\boldmath$\gamma^{(2)}_{\lambda}$}\\
        \hline
        $B$     &       4393    &       $-0.35$                         &       $15.6$  \\
        $R$     &       6470    &       $-0.24$                         &       $16.2$  \\
        $I$     &       8320    &       $-0.20$                         &       $16.9$  \\
        $J$     &       12660   &       $-0.113$                        &       $17.5$  \\
        $H$     &       16732   &       $-0.082$                        &       $17.9$  \\
        $K$     &       22152   &       $-0.045$                        &       $18.3$  \\
        \hline
\end{tabular}
\caption{\label{tab:gamma}
        Coefficients for the parameterisation of $\gamma_{\lambda}$ in each band.  Those for $B$, $R$, $I$ and $K$
        bands are taken from Tully \etal\ (1998). Coefficients for $J$ and $H$ bands are from
        interpolations described in the text.}
\end{table}

It should be noted that there may be some dependence of the $\gamma$ parameters upon the type of magnitude used.
Ideally these coefficients would be derived directly from the sample used in this analysis.
However, the statistics available in this sample are insufficient to improve on the above estimates.

\section{Applying the \tf\ relation} \label{sec:applying}

The galaxies in this sample have been chosen to be physically associated with
clusters, for the usual reason that within each cluster the galaxies can be assumed to be at the same distance.
The \tf\ relation for each cluster can therefore be constructed using apparent, rather than absolute, magnitudes
and hence the cluster's TF slope can be found independent of its distance.  This is obviously an advantage, as these
distances are not exactly determined, and indeed the aim is to use the derived relation to improve upon them!

However, each cluster has only a rather limited collection of spiral galaxies, and therefore the validity of the
relation suffers from low statistics.  In addition, the applicability of a relation derived solely from one cluster
to galaxies in other clusters, or the general field, is in question.
For the peculiar velocity field analysis intended in this project, we wish to determine the velocities with respect
to an average Hubble flow, for a number of distances and areas around the sky.
We therefore need a sample of many clusters, each containing enough galaxies to allow a reasonable independent TF
fit, but which may be combined to improve the statistics, and compared to determine the peculiar velocity field. 

\subsection{Bias consideration}
Before the procedure used in this project is described, we must first consider a number of possible biases
which may arise.
Even when fitting a TF relation to the data for each individual cluster, without direct dependence on distance,
a dependence may enter the relation indirectly, through selection effects and 
distance dependencies in the procedures used to `correct' the magnitudes and velocity widths.

The most obvious of these corrections is that for internal
extinction, described in \S \ref{sec:corintext}.  The $\gamma$ parameters of this 
correction (equation \ref{eqn:gammas}) contain a term involving the fully-corrected absolute magnitude.  
This seems rather contradictory,
since this magnitude is a desired product of the entire correction and calibration scheme, but the value of
its corresponding $\gamma$ parameter is required for calculating this magnitude in the first place.
An iterative approach therefore suggests itself.  The cosmological corrections (\S \ref{sec:corcos}) also
depend upon distance, through the redshift $z$, which should not actually be the measured redshift, but that
corresponding to the source's proper distance in a uniform Hubble expansion.  However, these are much smaller
corrections and therefore considered to have a negligible effect in comparison to internal extinction.
They are therefore calculated using values from the literature and not iterated over.

A more subtle source of bias, though at least as important, is introduced by the selection function.
This has been studied in detail \cite{Sandage88,Teerikorpi87}, and in particular by Willick \cite*{Willick94}.
The problem may be demonstrated by the following discussion.
Consider fitting the TF relation to some data using a least squares regression on the residuals in magnitude.
This is referred to as the `forward' method.  If the data sample is from a cluster, all the galaxies lie
at the same distance.  If the sample is complete, or limited in some way independent
to magnitude, then the fit will reproduce the `true' TF relation.  
The intrinsically brightest galaxies in the sample will tend to lie above this true correlation line. 
However if the sample is magnitude limited, which is usually the case, then at the `lower end' of the
relation these bright galaxies will be preferentially included over the fainter ones. 
Therfore the `lower end' of the fit will be `pulled up' to brighter magnitudes.  
The bias therefore tends decrease the TF slope, and moves the intercept to brighter magnitudes.
The more distant a group the worse its absolute magnitude completeness limit, and the higher up the relation
this bias sets in.
Therefore, comparing TF intercepts leads to more distant groups appearing closer than they would if a 
magnitude-complete sample were used. 
Unless this incompleteness bias is corrected for,
more distant groups will be found to lie closer than they actually are.  This is obviously a problem for
determining peculiar velocities relative to a Hubble flow.

An alternative, the `inverse' method --- in which the residuals in velocity width are minimised --- has been, and still
is, advocated as nullifying this bias \cite{TP2000}.  This is justified by claiming the magnitude limit cannot 
effect an `orthogonal' fit, and as long as there is no selection by velocity width, no bias is present.
In ideal circumstances the slope of the regression is independent of the magnitude limit \cite{Schechter80}.
However, the two quantities cannot be treated completely independently, and there may be subtle selection effects
on velocity width, especially from radio measurements.  It has been shown that bias is still present
in the inverse relation \cite{Willick94}, but this bias is around 5 times less than for the forward method.
G97II argues that the inverse regression does not completely nullify the bias in practice,
due to the large errors on measurements and especially corrections of the velocity widths.
They maintain that incompleteness bias  corrections are unavoidable, and implement a seemingly thorough, 
and reasonably complex, consideration and correction procedure.  
They favour a `bivariate' regression on both coordinates.

Tully \& Pierce \cite*{TP2000} still insist, however, that this is unnecessary and even unhelpful,
as it requires assumptions of the detailed properties of the population from which the sample is drawn.
These are not very well known.  They maintain that, for their analysis, the inverse regression method reduces the
bias to an acceptably small level.  Sakai \etal\ \cite*{Sakai_etal} also find, upon consideration of the correction
method of G97II, this cluster population incompleteness bias to be almost negligible, even for the forward method;
amounting to $\sim$$0.05$ mag, corresponding to $\sim$$2$\% in distance.

In this project the view is taken that, with care, the bias is negligible for
the inverse relation.
Ideally the analysis in this project could be performed for both forward and inverse methods, with and
without Willick's corrections, and the level of bias clearly assessed.  Unfortunately, time does not
permit this here.  However, a comparison with the results of G97I, G97II and G98s' fully-corrected method
in \S \ref{sec:results} helps to demonstrate that neglecting this bias is not as serious as some suggest. 

The effect known as Malmquist bias also deserves some mention.  Though often used when referring to the
cluster incompletness bias discussed above, this is a related but subtly different effect.  It arises
because, when observing in a small solid angle, there may be a different number of galaxies at a 
distance $r + \delta r$ than at $r$, even for a homogenous spatial distribution.
However, the problem is not relevant in this case.  The regions considered here are clusters,
and as such are very small (in depth and width) with respect to their distance.
Therfore the distance over which the number of galaxies, within a small solid angle, increases significantly is much
less than the cluster size.  The bias is therefore negligible for an individual cluster.
This bias is called cluster size-sample incompleteness bias in \S3.4.2 of G97II, where it is found to be small, 
$< 0.05$ mag for a very large cluster with size equal to $10$\% of its distance from us.

The clusters themselves may be susceptible to Malmquist bias.  However, the clusters are few in number,
sparsely distributed, and their selection includes reasons unrelated to distance.  The bias is therefore
very difficult to assess.  Fortunately, for the same reasons and the fact that the error on a cluster distance
is much smaller than that for a galaxy, this bias is very small.  G97II discuss this bias in \S3.5, where they
conclude it to be very small and do not apply any correction.  It is considered to be
negligible in this project.

\subsection{Fitting the TF}
All fits of the TF relation are done using the inverse method, with,
\begin{equation} \label{eqn:inverseTF}
\log{\Wcor} = a m^{k,b,i} + b
\end{equation}
where $m^{k,b,i}$ is the corrected apparent or absolute magnitude, depending upon the context.
This is fit by a least squares regression with weightings,
\begin{equation} \label{eqn:w_i}
w_i = \frac{1}{\sigma_{i,total}^2} = \frac{1}{\hat{\sigma}_{int}^2 + \sigma_{i,lw}^2}
\end{equation}
$\hat{\sigma}_{int}$ is constant for all points within a cluster, indicated by the hat, 
whereas the $\sigma_{i,lw}$ are the errors in line-width for each point.
Early studies ignored the intrinsic scatter in fitting, later studies used a constant value.
Modern studies \cite[for example]{Shellflow} include it as an independent free parameter in the fit.
G97II models the intrinsic scatter by fitting with ${\sigma}_{int} = 0$ and then fitting an intrinsic
scatter as a linear function of velocity width, to account for the excess scatter above that expected
from measurement and correction errors.
In the present analysis the amount of intrinsic scatter is estimated from the data 
for each cluster, to avoid introducing an assumed constant, but is assumed to be independent.  
The data is first fit without any weightings.  The rms
scatter from this fit, $\hat{\sigma}_{\circ}$, is then used as an estimate of the combined intrinsic scatter,
$\hat{\sigma}_{int}$, and average scatter due to errors on line-width, $\hat{\sigma}_{lw}$,
so that,
\begin{equation}
\hat{\sigma}_{int}^2 = \hat{\sigma}_{\circ}^2 - \hat{\sigma}_{lw}^2
\end{equation}
where $\hat{\sigma}_{lw}$ is derived from the $n$ individual $\sigma_{i,lw}$ in the standard way,
\begin{equation} \label{eqn:weights}
\hat{\sigma}_{lw}^2 = \frac{\sum_{i}^{n}{\sigma_{i,lw}^2}}{n}
\end{equation}
The main consequence of introducing an intrinsic scatter is to dampen the effect of those points with very small
velocity width errors, which otherwise over-dominate the regression.

It was realised that care must be taken when determining the errors on a weighted fit using intrinsic scatter,
as standard methods use the weightings to determine the measurement errors using equation \ref{eqn:weights}.
The errors on the fit then depend upon these measurement errors.
To avoid incorrect errors above their true values, the error calculation method has been adjusted to remove
a contribution from the intrinsic scatter proportion of the weighting.

\subsection{Algorithm implemented} \label{sec:algimp}

Each galaxy is treated as having the same distance as the cluster it is associated with, 
this distance initially being taken from table 1 of G97I assuming $H_0 = 75$ km s$^{-1}$ Mpc$^{-1}$.
They are then corrected for cosmological (\S \ref{sec:corcos}), Galactic (\S \ref{sec:corgalext}) 
and internal (\S \ref{sec:corintext}) extinction effects.

Firstly each cluster is fit in apparent magnitude to find individual TF slopes, $a_i$.
These are averaged to provide an initial estimate of the global TF slope, $a$.  
From this point on the same slope is used for every cluster.

The following is looped over until a stable solution is converged upon.
Each cluster is re-fit in apparent magnitude with the fixed slope, $a$, 
to produce TF intercepts for each cluster, $b_i$.
These intercepts are then used to calculate relative distances to each cluster, adjusted to approximately unity
by an arbitrary factor,
\begin{equation}
d_{relative} = 10^{-0.2(b_i/a + 25)}
\end{equation}  
These are then adjusted to absolute distances by assuming the Coma cluster, A1656, to be at a distance of
$7200$ km s$^{-1}$.

All the clusters are then fit together in absolute magnitude (derived from the just calculated distances),
to determine an improved estimate of $a$.
The RMS scatter of this fit is compared with that for the previous loop, and if the change is lower than
a threshold value then the iteration ends.  Otherwise, the internal extinction correction is re-calculated
using the new distances and the process repeats.

The iteration converges rapidly, with a change in RMS of less than 0.1\% after a few iterations.

\subsection{Tully-Fisher plots}
\begin{figure}[tb]
\centering
\subfigure{
        \label{fig:TFplotsEri:I}
        \includegraphics[width=200pt]{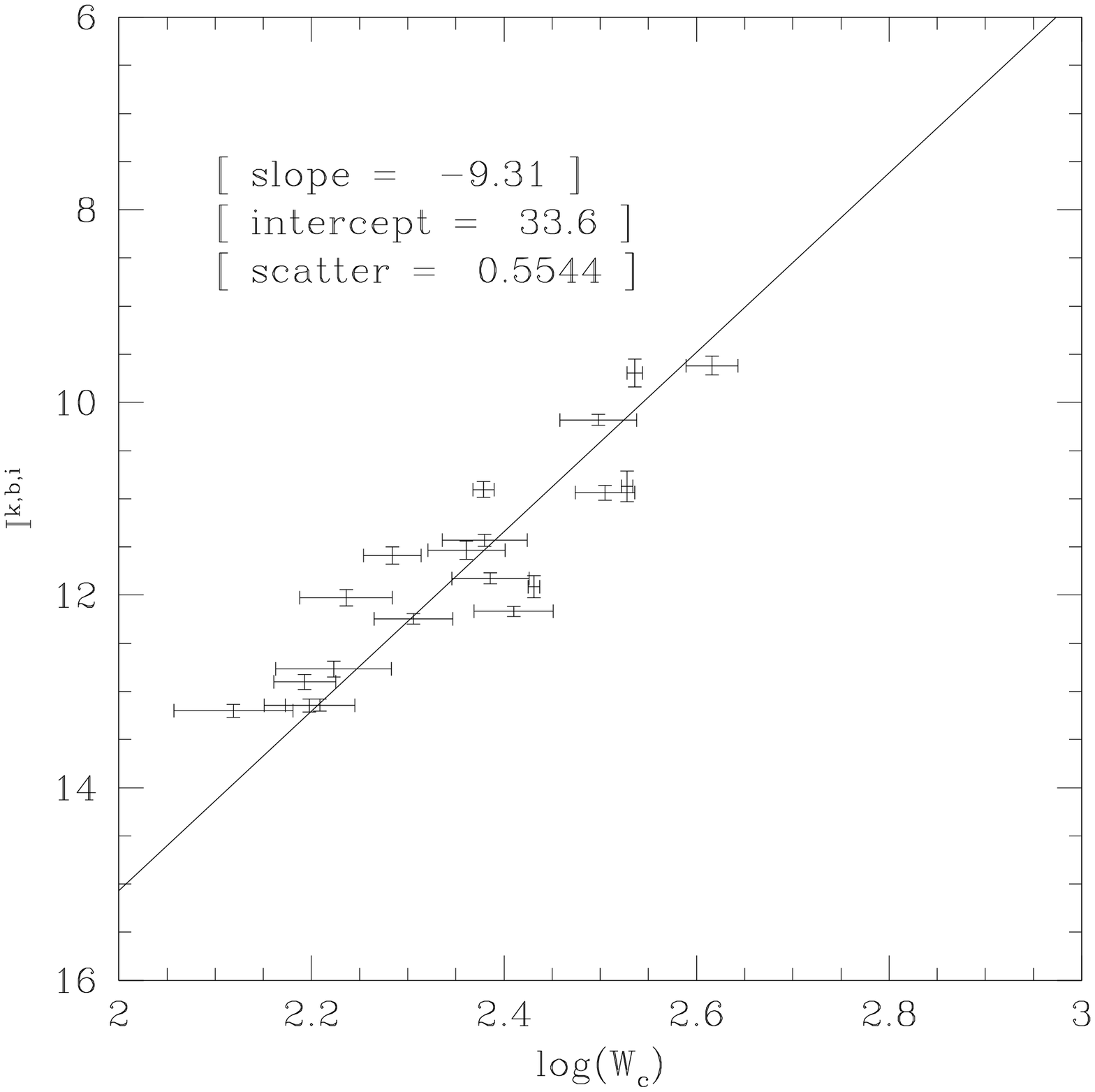}
}
\hfill
\subfigure{
        \label{fig:TFplotsEri:J}
        \includegraphics[width=200pt]{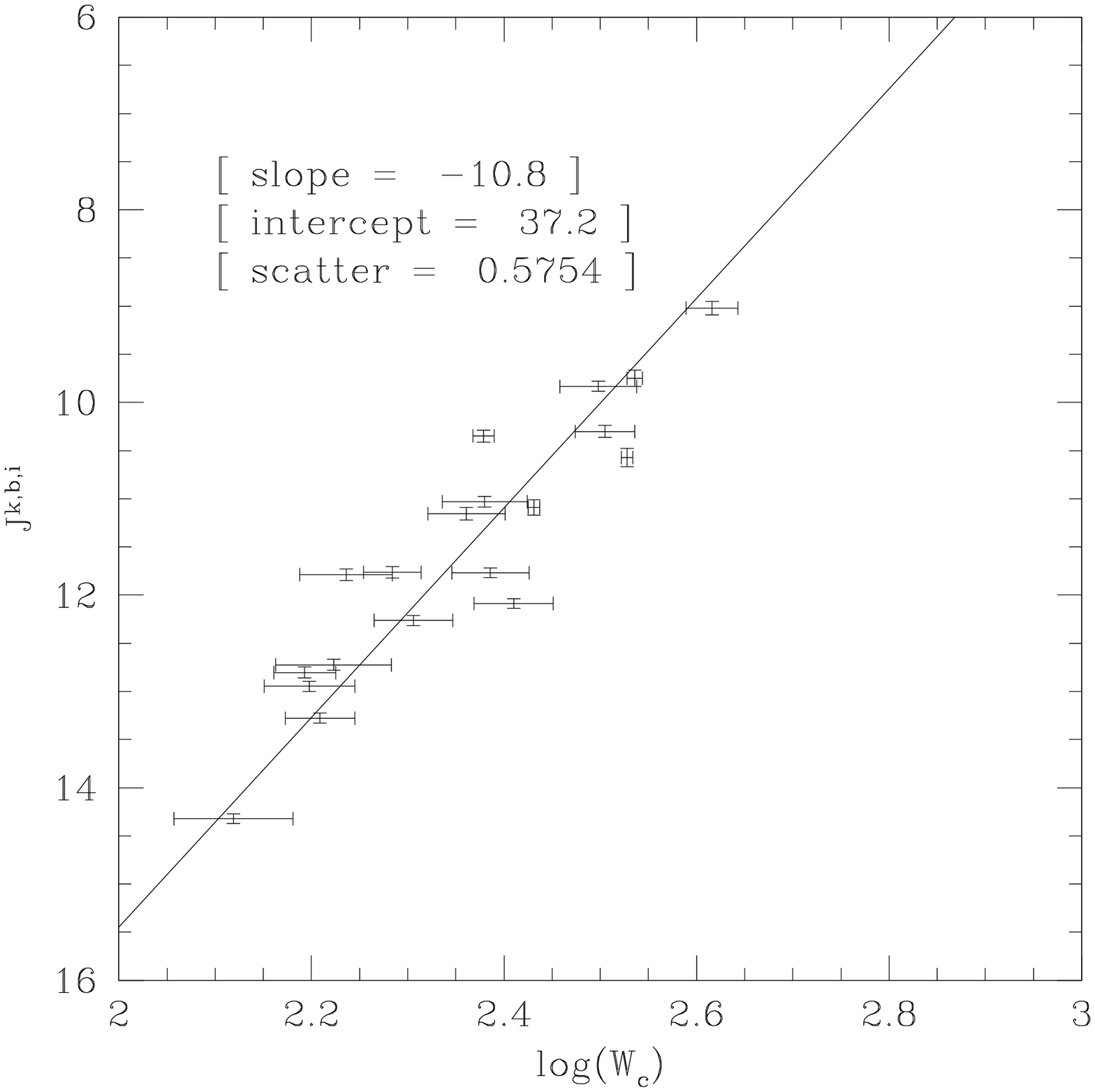}
}
\caption{\label{fig:TFplotsEri}TF plots for the Eridanus cluster. On the left is the $I$-band data,
and on the right is the $J$-band (k20fe) data. The $H$ and $K$-band plots are similar to $J$.
The line in each plot is the cluster fit for that band, as described in the text.}
\end{figure}

As an example of the individual cluster Tully-Fisher relations, the $I$ and $J$-band TF plots for Eridanus are 
shown in figure \ref{fig:TFplotsEri}.
The line on each plot is the corresponding TF fit for the cluster.  As specified above, in section \ref{sec:algimp},
the slope of this fit is the global value for that band, derived from fitting all the clusters, whereas the intercept 
has been found by fitting this slope to the cluster data. Both these parameters, and the scatter, have been 
approximately converted to forward fit values, to ease comparison. The quoted scatter is therefore in magnitudes.

\begin{figure}[tb]
\centering
\subfigure{
        \label{fig:TFplotsAll:I}
        \includegraphics[width=200pt]{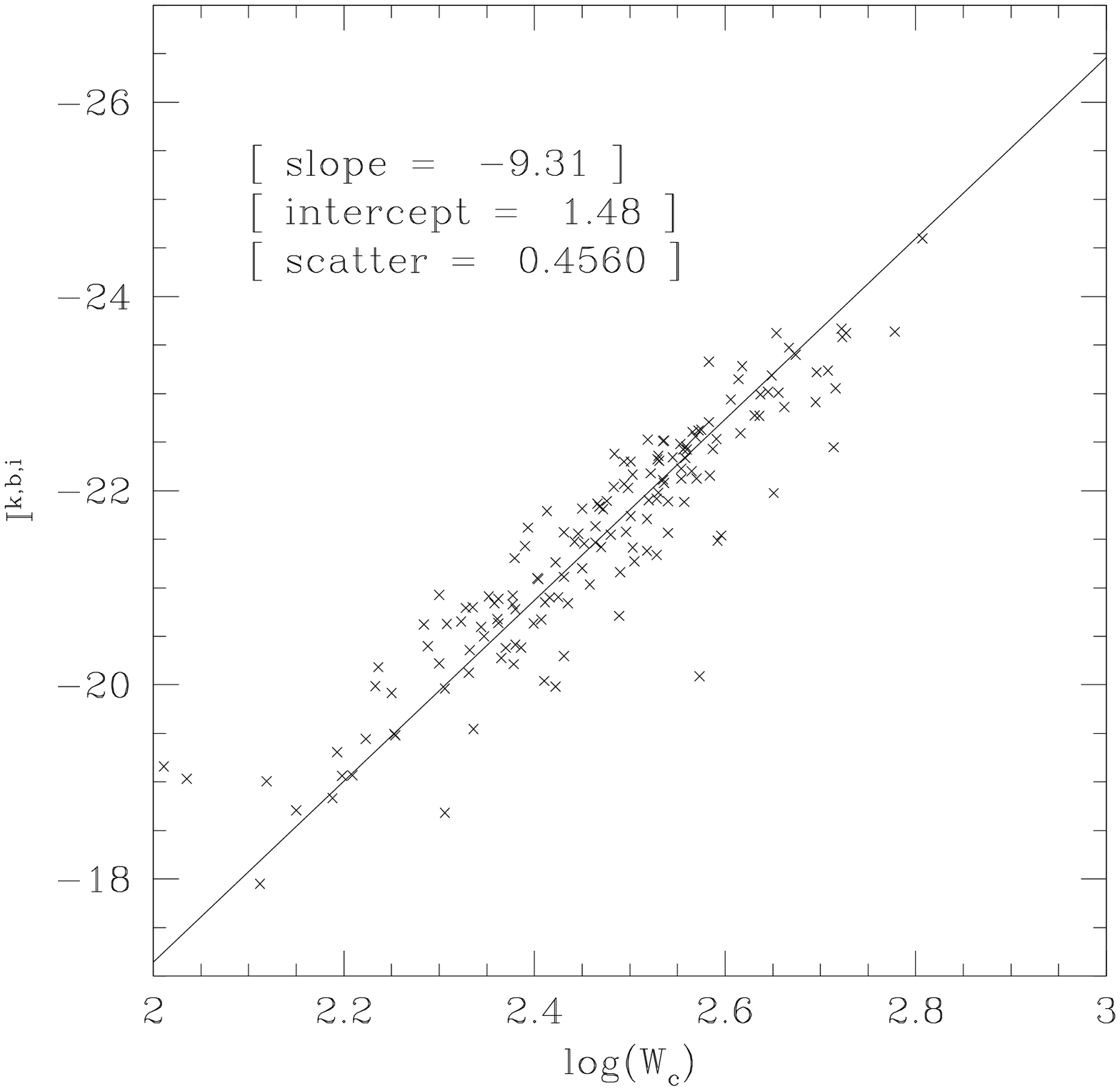}
}
\hfill
\subfigure{
        \label{fig:TFplotsAll:J}
        \includegraphics[width=200pt]{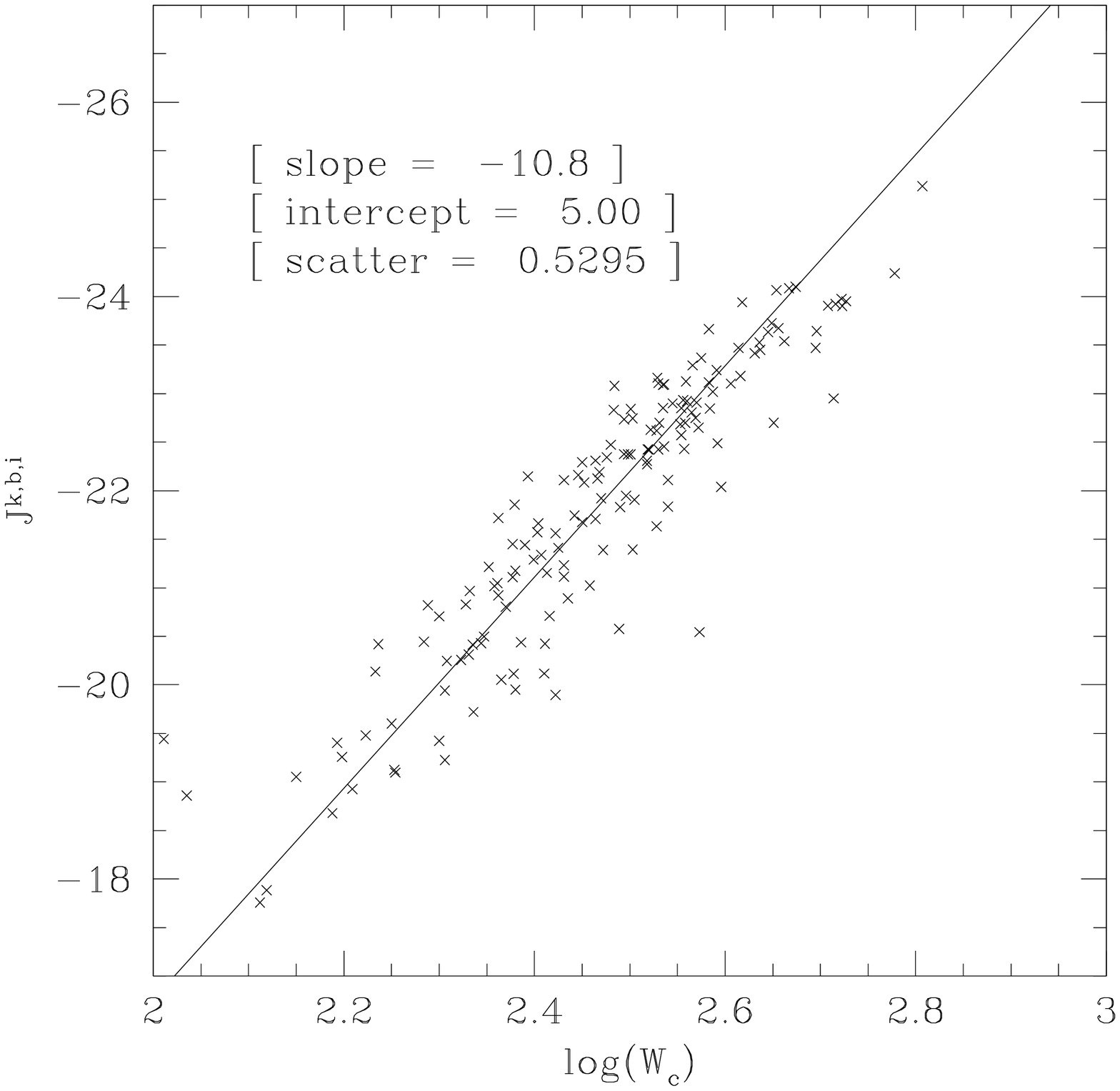}
}
\caption{\label{fig:TFplotsAll}TF plots for all the sample clusters combined in absolute magnitude. 
On the left is the $I$-band data, and on the right is $J$-band (k20fe) data. 
The $H$ and $K$-band plots are similar to $J$.
The line in each plot is the fit to the data, as described in the text.}
\end{figure}

TF plots of all the clusters together, combined using the prescription of section \ref{sec:algimp},
are shown in figure \ref{fig:TFplotsAll}. Error bars are omitted for clarity. The clusters are combined in 
absolute magnitude, using the distances resulting from the iterative process described in section \ref{sec:sample}.
The fits to the data in each plot are shown.

The data used in all these plots are from the final sample, as discussed in section \ref{sec:sample}.

Note that the scatter for the $I$-band G97 data and the $J$-band 2MASS data are comparable.
The $H$ and $K$-band data give scatter similar to $J$. 
Given the larger photometry errors for 2MASS this indicates similar intrinsic scatter
for all these bands.

\section{Measured distances and peculiar velocities} \label{sec:results}
The global TF slopes for each band are given in table \ref{tab:TFslopes}, and the TF intercepts
and scatter, for each band and each cluster, in table \ref{tab:TFfits}

The `inverse' values given are those directly from the fits.  The `forward' slope quoted is the reciprocal
of the `inverse' slope, and is not equal to the slope that would result from a forward fit, as noted
by Willick \cite*{Willick94}.  The `forward' scatter is similarly the `inverse' scatter divided by the
modulus of the slope, and is a representation of the scatter in magnitudes.  The `intrinsic' scatter values
are the result of subtracting in quadrature the magnitude errors from the forward scatter.  This is not
the true intrinsic scatter as the line-width errors are still included.  Also, the raw magnitude errors
have not been properly incorporated, but assumed to be a constant value of 0.05 mag.  This is lower than
the true 2MASS errors, and hence the true intrinsic magnitudes in $J$, $H$ and $K$ are lower and more
comparable with $I$ than those listed.

The results of the TF fits described above are listed, for each band, in table \ref{tab:TFresults}.
The distances are given in terms of a redshift relative to Coma A1656 at $7200$ km s$^{-1}$.
Also included are the average redshift for the cluster, and the peculiar velocity calculated
as the difference between this and the distance averaged over the bands.

\begin{figure}[p]
\centering
\includegraphics[width=300pt]{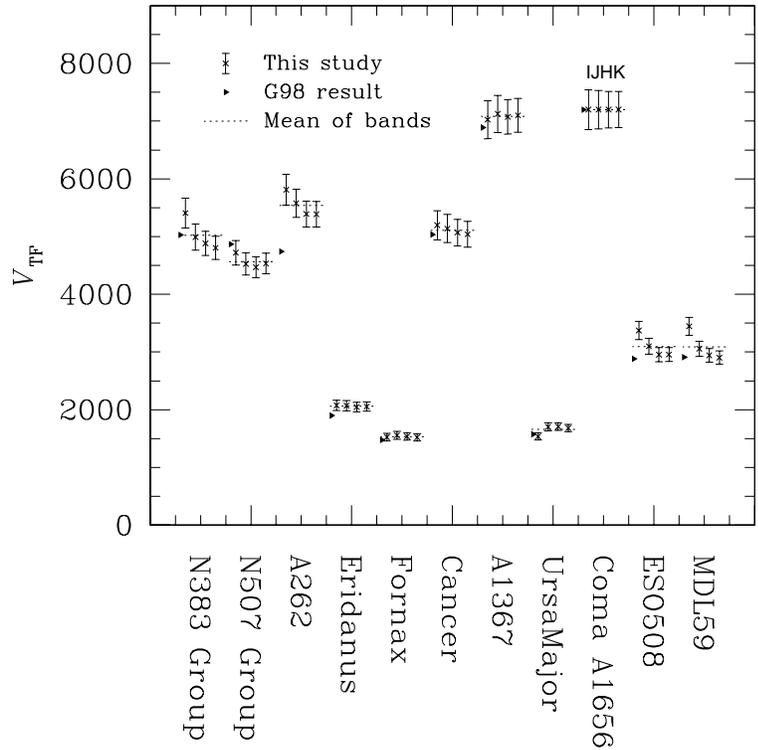}
\caption{\label{fig:dist}
A plot of the cluster distances measured in this project, for each band, with the values measured
by G98 for comparison.}
\end{figure}

The cluster distances are plotted in figure \ref{fig:dist}.  The distances derived from the 
different bands generally agree well, especially the 2MASS bands, but also $I$.  The 2MASS distances
do tend to lie below the $I$-band derived distances, however.  The distances also
mostly agree with the distances found by G98 within the errors.  However, some clusters differ
by $\gtrsim 1 \sigma$, and especially A262 differs significantly.
No reasons have been found for this, and the effect appears to be due mostly to the difference in statistics
between the studies.

\begin{figure}[p]
\centering
\includegraphics[width=220pt]{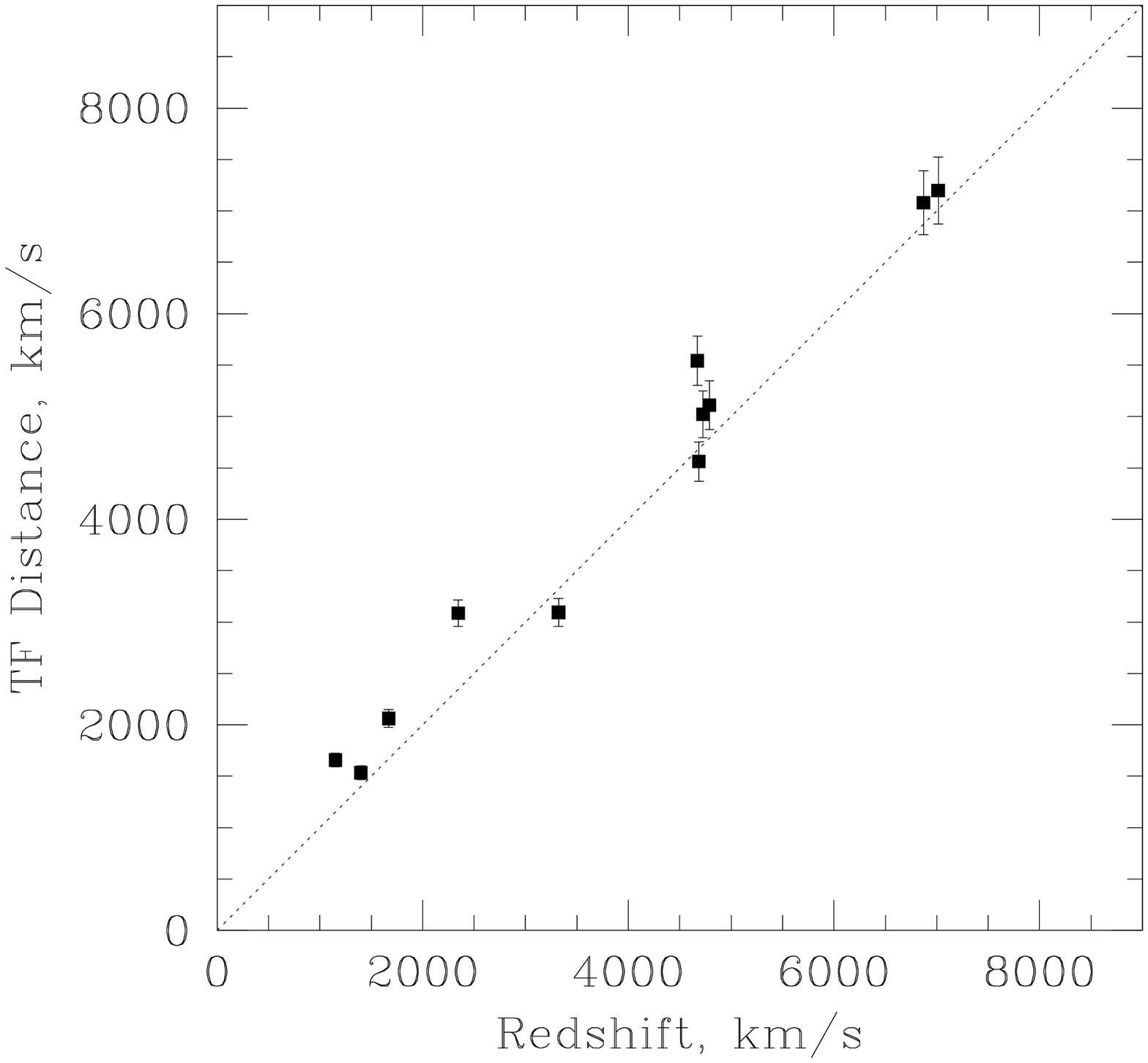}
\caption{\label{fig:hubble}
The Hubble diagram for the cluster distances measured in this project, plotted against their redshifts.}
\end{figure}

Combining the distances with cluster
redshifts (calculated as the average of the sample galaxies in the cluster) 
gives the Hubble diagram in figure \ref{fig:hubble}.  It can be seen in this figure that a number
of points are significantly removed from the mean Hubble flow, implying peculiar velocities,
tabulated in table \ref{tab:TFresults}.
The absolute calibration was arbitrarily set by putting Coma A1656 at $7200$ km s$^{-1}$, and so 
this may be freely varied slightly.  This is equivalent to moving the points in figure \ref{fig:hubble}
horizontally by a constant amount.  This would give more agreement with the Hubble flow, but
peculiar velocities would remain.  These are in some agreement with those derived by G98,
to the same degree as the distances.  However, as before, A262 is in particular disagreement.

In the late stages of this project, the analysis has been repeated using Kron circular fiducial
aperture magnitudes.  The results are the same, with cluster distances varying only by few percent,
mostly upwards to agree slightly better with the $I$-band derived distances.

\begin{table}[tb]
\centering
\begin{tabular}{cccccc}
        \hline
        \textbf{Band} & {\boldmath $a_{\text{\bf inverse}}$} &
                        {\boldmath $\sigma_{\text{\bf total inverse}}$} &
                        {\boldmath $a_{\text{\bf forward}}$} &
                        {\boldmath $\sigma_{\text{\bf total forward}}$} {\scr (mag)} & 
                        {\boldmath $\sigma_{\text{\bf intrinsic}}$} {\scr (mag)}\\
        \hline
        $I$             & $-0.1073$     & $0.0490$      & $-9.32$       & $0.456$       & $0.375$ \\
        $J$             & $-0.0919$     & $0.0487$      & $-10.87$      & $0.530$       & $0.474$ \\
        $H$             & $-0.0916$     & $0.0483$      & $-10.92$      & $0.527$       & $0.478$ \\
        $K$             & $-0.0901$     & $0.0494$      & $-11.09$      & $0.548$       & $0.504$ \\
        \hline
\end{tabular}
\caption{\label{tab:TFslopes} Global TF slopes derived from the final absolute magnitude TF fit of \S \ref{sec:algimp}.}
\end{table}

\begin{table}[p]
\centering \footnotesize
\begin{tabular}{lc@{\hspace{25pt}}c@{\hspace{25pt}}c}
        \hline
        \textbf{Cluster} & \textbf{Band} & {\boldmath $b_{\text{\bf inverse}}$} & 
                {\boldmath $\sigma_{\text{\bf total}}$} {\scr (mag)}\\
        \hline
        N 383 Group (Pisces)            & I     & $3.841$       & $0.552$       \\
                                        & J     & $3.596$       & $0.651$       \\
                                        & H     & $3.530$       & $0.609$       \\
                                        & K     & $3.493$       & $0.431$       \\
        \hline
        N 507 Group                     & I     & $3.809$       & $0.364$       \\
                                        & J     & $3.576$       & $0.272$       \\
                                        & H     & $3.513$       & $0.300$       \\
                                        & K     & $3.482$       & $0.324$       \\
        \hline
        A 262                           & I     & $3.858$       & $0.337$       \\
                                        & J     & $3.618$       & $0.408$       \\
                                        & H     & $3.550$       & $0.445$       \\
                                        & K     & $3.516$       & $0.455$       \\
        \hline
        Eridanus                        & I     & $3.618$       & $0.554$       \\
                                        & J     & $3.420$       & $0.575$       \\
                                        & H     & $3.357$       & $0.572$       \\
                                        & K     & $3.327$       & $0.561$       \\
        \hline
        Fornax (S 0373)                 & I     & $3.546$       & $1.083$       \\
                                        & J     & $3.363$       & $1.247$       \\
                                        & H     & $3.300$       & $1.213$       \\
                                        & K     & $3.268$       & $1.267$       \\
        \hline
        Cancer                          & I     & $3.832$       & $0.427$       \\
                                        & J     & $3.602$       & $0.514$       \\
                                        & H     & $3.538$       & $0.560$       \\
                                        & K     & $3.503$       & $0.580$       \\
        \hline
        A 1367                          & I     & $3.902$       & $0.378$       \\
                                        & J     & $3.667$       & $0.431$       \\
                                        & H     & $3.604$       & $0.313$       \\
                                        & K     & $3.570$       & $0.300$       \\
        \hline
        Ursa Major                      & I     & $3.548$       & $0.542$       \\
                                        & J     & $3.381$       & $0.512$       \\
                                        & H     & $3.321$       & $0.498$       \\
                                        & K     & $3.288$       & $0.517$       \\
        \hline
        Coma (A 1656)                   & I     & $3.908$       & $0.340$       \\
                                        & J     & $3.669$       & $0.492$       \\
                                        & H     & $3.607$       & $0.501$       \\
                                        & K     & $3.573$       & $0.526$       \\
        \hline
        ESO 508                         & I     & $3.731$       & $0.378$       \\
                                        & J     & $3.501$       & $0.335$       \\
                                        & H     & $3.430$       & $0.309$       \\
                                        & K     & $3.398$       & $0.316$       \\
        \hline
        MDL 59                          & I     & $3.736$       & $0.442$       \\
                                        & J     & $3.498$       & $0.658$       \\
                                        & H     & $3.429$       & $0.632$       \\
                                        & K     & $3.395$       & $0.731$       \\
        \hline
\end{tabular}
\caption{\label{tab:TFfits}
TF inverse fit intercept, $b_{\text{inverse}}$, 
total scatter, $\sigma_{\text{total}}$, and intrinsic scatter, $\sigma_{\text{intrinsic}}$,
derived from the TF fits of \S \ref{sec:algimp}.}
\end{table}

\begin{table}[p]
\centering \footnotesize
\begin{tabular}{lc@{\hspace{20pt}}r@{\scr $\ \pm\ $}l@{\hspace{25pt}}c@{\hspace{25pt}}c}
        \hline
        \textbf{Cluster} & \textbf{Band} & \multicolumn{2}{l}{\boldmath $V_{\text{\bf TF}}$ {\scr km s$^{-1}$}} & 
                {\boldmath $\overline{cz}$ {\scr km s$^{-1}$}} & 
                {\boldmath $\overline{V}_{\text{\bf peculiar}}$ {\scr km s$^{-1}$}}\\
        \hline
        N 383 Group (Pisces)            & I     & $5407$        & $258$         & $4725$        & $-296$ \\
                                        & J     & $4990$        & $228$         \\
                                        & H     & $4882$        & $210$         \\
                                        & K     & $4805$        & $203$         \\
        \hline
        N 507 Group                     & I     & $4720$        & $213$         & $4687$        & $125$ \\
                                        & J     & $4527$        & $191$         \\
                                        & H     & $4469$        & $178$         \\
                                        & K     & $4534$        & $178$         \\
        \hline
        A 262                           & I     & $5812$        & $269$         & $4672$        & $-869$ \\
                                        & J     & $5576$        & $246$         \\
                                        & H     & $5391$        & $224$         \\
                                        & K     & $5386$        & $220$         \\
        \hline
        Eridanus                        & I     & $2075$        & $91$          & $1669$        & $-393$ \\
                                        & J     & $2071$        & $88$          \\
                                        & H     & $2048$        & $82$          \\
                                        & K     & $2054$        & $81$          \\
        \hline
        Fornax (S 0373)                 & I     & $1526$        & $67$          & $1400$        & $-135$ \\
                                        & J     & $1556$        & $68$          \\
                                        & H     & $1537$        & $64$          \\
                                        & K     & $1523$        & $62$          \\
        \hline
        Cancer                          & I     & $5196$        & $253$         & $4790$        & $-321$ \\
                                        & J     & $5138$        & $244$         \\
                                        & H     & $5069$        & $228$         \\
                                        & K     & $5039$        & $223$         \\
        \hline
        A 1367                          & I     & $7025$        & $329$         & $6871$        & $-210$ \\
                                        & J     & $7125$        & $321$         \\
                                        & H     & $7075$        & $299$         \\
                                        & K     & $7100$        & $294$         \\
        \hline
        Ursa Major                      & I     & $1537$        & $63$          & $1152$        & $-506$ \\
                                        & J     & $1703$        & $70$          \\
                                        & H     & $1708$        & $66$          \\
                                        & K     & $1684$        & $14$          \\
        \hline
        Coma (A 1656)                   & I     & $7200$        & $344$         & $7014$        & $-186$ \\
                                        & J     & $7200$        & $333$         \\
                                        & H     & $7200$        & $316$         \\
                                        & K     & $7200$        & $311$         \\
        \hline
        ESO 508                         & I     & $3373$        & $158$         & $3322$        & $227$ \\
                                        & J     & $3099$        & $137$         \\
                                        & H     & $2953$        & $122$         \\
                                        & K     & $2955$        & $120$         \\
        \hline
        MDL 59                          & I     & $3445$        & $154$         & $2346$        & $-741$ \\
                                        & J     & $3056$        & $131$         \\
                                        & H     & $2942$        & $118$         \\
                                        & K     & $2903$        & $114$         \\
        \hline
\end{tabular}
\caption{\label{tab:TFresults}
Cluster distances (expressed as velocities), $V_{\text{TF}}$, average redshifts, $\overline{cz}$,
and peculiar velocities, $\overline{V}_{\text{peculiar}}$, derived from the TF fits of \S \ref{sec:algimp}.}
\end{table}

\section{Conclusions \& prospects for further work} \label{sec:conclusions}
The results of combining 2MASS photometry with velocity width data from the SCI sample of G97I
agree well with the same process applied with the $I$-band photometry of G97I.  This
indicates that 2MASS photometry is of a high enough quality to for use with the TF, while having
the advantage of exceptional all-sky homogeneity.  That comparable results may be acheived by using
2MASS magnitudes as by taking dedicated photometry, of the type presented by G97I, is a useful
finding.

Comparison with the results of the G97II and G98
analysis of the SCI sample shows reasonable agreement.  The correction procedure in this project, though
quite different to G97II, appears to have a roughly equivalent effect.  There are some significant 
disagreements between cluster distances, particularly for A262.  This issue is unresolved, but appears to be
caused by the much more limited statistics of this study, 
in turn due to the limited 2MASS coverage available at this time.

Two 2MASS magnitudes have been considered, both with different advantages for TF work.
These are the $K$-fiducial 20th mag arcsec$^{-1}$ isophotal elliptical aperture and the
$K$-fiducial Kron circular aperture.  Both give consistent results, implying the analysis is
not overly sensitive to the precise nature of the magnitude used.

This study suggests that applying the TF relation with 2MASS data gives useful results,
and that a full study with the complete 2MASS final release would be very helpful in verifying and 
building upon recent studies of the local velocity field.

\section*{Acknowledgements}
Many thanks to my supervisor, Dr. John Lucey, for guiding me through this project,
and forcing me to keep improving my understand of it.
Thanks to my friends for useful discussions, distractions and games of pool,
and to Cally Kennett, for her encouragement.
  
I am very grateful to Giovanelli and collaborators, for making their data compilation available publicly,
and to the 2MASS team for their excellent XSC catalogue.

This publication makes use of data products from the Two Micron All Sky Survey, 
which is a joint project of the University of Massachusetts and the 
Infrared Processing and Analysis Center/California Institute of Technology, 
funded by the National Aeronautics and Space Administration and the National Science Foundation.

This research has made use of the NASA/IPAC Extragalactic Database (NED) which is operated by the 
Jet Propulsion Laboratory, California Institute of Technology, 
under contract with the National Aeronautics and Space Administration.

\begin{singlespace}
{
\small
\bibliographystyle{astron}
\bibliography{mnemonic,thesis}
}
\end{singlespace}

\end{document}